\documentclass[fleqn,usenatbib]{mnras}

\usepackage{newtxtext,newtxmath}

\usepackage[T1]{fontenc}
\usepackage{ae,aecompl}
\usepackage[dvipsnames]{xcolor}
\usepackage{newtxtext,newtxmath}

\usepackage[T1]{fontenc}
\usepackage{ae,aecompl}

\usepackage{graphicx}	
\usepackage{amsmath}	

\title[N-rich stars in the Galaxy]{The contribution of N-rich stars to the Galactic stellar halo using APOGEE red giants}

\author[D. Horta et al.]{
Danny Horta$^{1}$\thanks{E-mail: 
D.HortaDarrington@2018.ljmu.ac.uk},
J. Ted Mackereth$^{2,3}$,
Ricardo P. Schiavon$^{1}$,
Sten Hasselquist$^{4,5}$,
\newauthor
Jo Bovy$^{2,3}$,
Carlos Allende Prieto$^{6,7}$,
Timothy C. Beers$^{8}$,
Katia Cunha$^{9,10}$,
\newauthor
D. A. Garc\'ia-Hern\'andez$^{6,7}$,
Shobhit S. Kisku$^{1}$,
Richard R. Lane$^{11,12}$,
\newauthor
Steven R. Majewski$^{13}$,
Andrew C. Mason$^{1}$,
David M. Nataf$^{14}$,
\newauthor
Alexandre Roman-Lopes$^{15}$,
Mathias Schultheis$^{16}$
\\
$^{1}$Astrophysics Research Institute, Liverpool John Moores University, 146 Brownlow Hill, Liverpool L3 5RF, UK \\
$^{2}$ Canadian Institute for Theoretical Astrophysics, University of Toronto, 60 St. George Street, Toronto, ON, M5S 3H8, Canada \\
$^{3}$ Dunlap Institute for Astronomy and Astrophysics, University of Toronto, 50 St. George Street, Toronto, ON M5S 3H4, Canada \\
$^{4}$New Mexico State University, Las Cruces, NM 88003, USA\\
$^{5}$Department of Physics $\&$ Astronomy, University of Utah, Salt Lake City, UT 84112, USA\\
$^{6}$ Instituto de Astrof\'isica de Canarias, E-38200 La Laguna,Tenerife,
Spain \\
$^{7}$Universidad de La Laguna (ULL), Departamento de Astrof\'isica, Universidad de La Laguna, E-38206
La Laguna, Tenerife, Spain \\
$^{8}$Department of Physics and JINA Center for the Evolution of the Elements, University of Notre Dame, Notre Dame, IN 46556, USA\\
$^{9}$University of Arizona, Tucson, AZ 85719, USA\\
$^{10}$Observat\'orio Nacional, S\~ao Crist\'ov\~ao, Rio de Janeiro, Brazil\\
$^{11}${Instituto de Astronom\'ia y Ciencias Planetarias, Universidad de
Atacama, Copayapu 485, Copiap\'o, Chile}\\
$^{12}$ Instituto de Astrof\'{i}sica, Pontificia Universidad Cat\'{o}lica de Chile, Av. Vicuna Mackenna 4860, 782-0436 Macul, Santiago, Chile\\
$^{13}$ Dept. of Astronomy, University of Virginia, Charlottesville, VA 22904-4325, USA\\
$^{14}$Center for Astrophysical Sciences and Department of Physics and Astronomy,
The Johns Hopkins University,
Baltimore, MD 21218\\
$^{15}$Departamento de Astronom\'ia, Universidad de La Serena, Cisternas 1200, La Serena, Chile\\
$^{16}$ Universit\'e C\^ote d'Azur, Observatoire de la C\^ote d'Azur, 
Laboratoire Lagrange, CNRS, Blvd de l'Observatoire, F-06304 Nice, France \\
}

\date{Accepted XXX. Received YYY; in original form ZZZ}

\pubyear{2020}
\begin{document}
\label{firstpage}
\pagerange{\pageref{firstpage}--\pageref{lastpage}}
\maketitle
\begin{abstract}

The contribution of dissolved globular clusters (GCs) to the stellar content of the Galactic halo is a key constraint on models for GC formation and destruction, and the mass assembly history of the Milky Way. Earlier results from APOGEE pointed to a large contribution of destroyed GCs to the stellar content of the inner halo, by as much as 25$\%$, which is an order of magnitude larger than previous estimates for more distant regions of the halo. We set out to measure the ratio between N-rich and normal halo field stars, as a function of distance, by performing density modelling of halo field populations in APOGEE DR16. Our results show that at 1.5 kpc from the Galactic Centre, N-rich stars contribute a much higher 16.8$^{+10.0}_{-7.0}$$\%$ fraction to the total stellar halo mass budget than the 2.7$^{+1.0}_{-0.8}$$\%$ ratio contributed at 10 kpc. Under the assumption that N-rich stars are former GC members that now reside in the stellar halo field, and assuming the ratio between first-and second-population GC stars being 1:2, we estimate a total contribution from disrupted GC stars of the order of 27.5$^{+15.4}_{-11.5}$$\%$ at r = 1.5 kpc and 4.2$^{+1.5}_{-1.3}$$\%$ at r = 10 kpc. Furthermore, since our methodology requires fitting a density model to the stellar halo, we integrate such density within a spherical shell from 1.5-15 kpc in radius, and find a total stellar mass arising from dissolved and/or evaporated GCs of $M_{\mathrm{GC,total}}$ = 9.6$^{+4.0}_{-2.6}$ $\times$ 10$^{7}$ M$\odot$.
\end{abstract}


\section{Introduction}
It is common knowledge that the globular cluster (GC) system of the Milky Way holds crucial insights into the formation and assembly history of the Galaxy. A milestone in this field was the work by \citet{Searle1978}, who used the Galactic GC population to infer that GCs found in the outer Galactic halo region formed over a longer period than those found in the inner-halo region, leading to the conclusion that the former population originated from accreted satellite systems. This concept has since been refined with improved measurements of ages and chemical abundances, leading to the discovery of a bifurcation in the age-metallicity distribution of Galactic GCs \citep{Marin-Franch2009}, where the young and metal-poor branch traces the population of GCs that are thought to result from satellite galaxy accretion (\citealp[e.g.,][]{Forbes2010,Leaman2013}). Similarly, this division is also reflected in the $\alpha$ and Fe compositions of GCs with metallicities between --1.5 < [Fe/H] < --1, whereby accreted GCs typically present lower [$\alpha$/Fe] for a given [Fe/H] in comparison to their \emph{in situ} counterparts \citep[e.g.,][]{Horta2020}. The results obtained from these observational studies fit well in the present cosmological framework, the $\Lambda$-CDM model (\citealp[e.g.,][]{Schaeffer1985,Frenk2012}). In this cosmological paradigm, larger galaxies engulf smaller satellite galaxies and grow through a process of hierarchical mass assembly. 
As a result, a substantial fraction of the GCs from those accreted galaxies may survive the merger, depending on the details of the accretion, as suggested by various observational results (\citealp[e.g.,][]{Brodie2006,Myeong2019,Massari2019,Koppelman2019,Kruijssen2019,Forbes2020}).

To elucidate the role GCs play in the formation and mass assembly of their host galaxies, it is vital to first understand how GCs form and evolve in a cosmological context. The leading GC formation scenarios propose a framework in which GCs formed in the turbulent discs of their host galaxies at \textit{z} $\sim$2-3, where, due to tidal shocks (\citealp{Gnedin2001}) and the so called "cruel cradle effect" (\citealp{Kruijssen2012b}), GCs formed \textit{in situ} were largely destroyed (\citealp{Elmegreen2010,Kruijssen2011,Kruijssen2014,Kruijssen2015}). As galaxies evolved, mergers redistributed GC systems of accreted satellites onto the host galaxies, typically depositing them in the outer regions of the stellar halo component, where mass loss via evaporation takes place in a longer timescale.

The question of how much stellar mass in the halo arises from GC dissolution and/or evaporation emerges as a natural implication from the above theoretical framework. In an attempt to answer such a question, over the last decade there has been substantial work focusing on the identification of GC stars that have been stripped from their parent GC, and now reside in the halo field of the Galaxy (\citealp[e.g.,][Kisku et al., in prep]{Martell2010,Martell2016,Schiavon2017,Koch2019,Hanke2020}). In these studies, stars arising from GC dissolution and/or evaporation were identified either by selecting field stars that present high nitrogen and low carbon abundances, or by identifying stars with strong CN bands, a feature that implies a higher N content and one that is typically found in the so called "second population" (SP) GC stars \citep[for a full review, see][]{Bastian2018}. Such studies have focused on both the outer regions of the stellar halo (\citealp[][]{Martell2010,Martell2016,Koch2019}), and within the inner few kiloparsecs from the Galactic Centre (\citealp{Schiavon2017}). Specifically, \citet{Martell2016} focused on metal poor ([Fe/H] $\leq$ --1.3) sample from the twelfth APOGEE data release and identified halo field stars that present high [N/Fe] and [Al/Fe] values, thus selecting former SP GC escapee candidates. Using the number ratio of N-rich stars to halo field stars, \citet{Martell2016} obtained an estimate for the contribution from dissolved SP GC stars to the outer regions of the stellar halo to be on the order of 2-3$\%$. A similar estimate was also obtained in a more recent study using the SEGUE data \citep{Yanny2009} by \citet{Koch2019}, which found a value of$\sim$2.6$\%$. Separately, \citet{Schiavon2017} analysed APOGEE data for stellar populations within $\sim$3 kpc of the Galactic Centre and found a much larger fraction of N-rich stars than their outer halo counterpart. In their study, \citet{Schiavon2017} identified field stars that presented high [N/Fe] and low [C/Fe] compared to the main population at a given metallicity, and determined a minimum contribution to the stellar halo from dissolved and/or evaporated SP GC stars of approximately$\sim$13-17$\%$. This results in a much higher contribution rate than in the outer halo, by a factor of$\sim$4-5.

Furthermore, recent theoretical studies making use of the suite of E-MOSAICS hydrodynamical cosmological simulations (\citealp{Pfeffer2018,Kruijssen2019_emosaics}) have assessed the contribution of stars dissolved from GCs to the stellar halo field for $\sim$25 Milky Way analogues. The results from these works predict a smaller contribution from SP GC stars than observational estimates, predicting a total contribution of the order of $\sim$1$\%$ \citep{Reina2019} for the outer halo region and $\sim$0.2-9$\%$ for the inner Galaxy \citep{Hughes2020}. Although the predicted fractions at face value do not match that of the observed fractions, an inner-to-outer halo ratio discrepancy is still predicted.

In connection with the mass contribution of dissolved GC stars to the Galactic stellar halo, the question of the spatial distribution of GC escapees is also an important one, and similarly, its stellar density as a function of Galactocentric distance. Answering these questions is non-trivial due to the difficulty in correcting the observed number of a tracer population to the total underlying sample. Along the same lines, estimates of the stellar mass are rigidly connected to the derived mass normalisation per tracer star, which generally relies on intricate calibration and stellar models. Moreover, GCs are typically associated to the stellar halo component of the Galaxy, and with no previous knowledge of the density profile of GC escapees, it is important to first understand how the stellar halo is spatially distributed. This information can be used as a guide to understand the density profile of GC-escapee stars, and will serve later for assessing the contribution of N-rich stars to the stellar halo field.

Recently, \citet{Iorio2018} modelled the density of the stellar halo using a cross-matched sample of \textit{Gaia}-2MASS RR Lyrae stars. In that work, several density profiles ranging in parameter complexity were tested, and it was found that the stellar halo was best fit by a triaxial ellipsoid, rotated with respect to the Galactic reference frame (in which the $X$ axis connects the Sun to the Galactic Centre) by $\sim$ 70$^{\circ}$; such rotation was thought to be introduced by a single massive merger event. Using Red Giant Branch (RGB) stars from the APOGEE fourteenth data release \citet{Mackereth2019b} also found this to be the case. The same study estimated a total halo stellar mass on the order of $\sim$1.3 $\times$ 10$^{9}$ M$\odot$ when considering the halo as an assembly of many individual components in the [Mg/Fe]-[Fe/H] plane. This result is in agreement with the recent work of \citet{Deason2019}, based on RGB number counts and extrapolation of the \citet{Einasto1965} profile, which also suggests a higher mass for the stellar halo of $\sim$1.4 $\times$ 10$^{9}$ M $\odot$. These works utilised existing density models to understand the shape and density profile of the stellar halo component of the Galaxy. Using such density modelling procedures, they were able to estimate the total mass of the stellar halo within a given Galactocentric radius. Such methods can be applied in theory to any tracer stellar population, enabling the estimation of mass as a function of Galactocentric radius while accounting for selection effects induced by observational data.

In this paper, we build on the previous work by \citet{Mackereth2019b}, and present a density profile, mass estimate, and a percentage contribution to the stellar halo from likely SP GC (i.e. N-rich) stars arising from GC dissolution and/or evaporation as a function of Galactocentric distance. This is accomplished via density modelling of APOGEE red giant stars, selected on the basis of their chemical compositions. Our methodology allows for the assessment of the stellar mass contained in the complete halo and the N-rich samples, as well as their corresponding mass as a function of Galactocentric distance. As explained in \citet{Mackereth2019b}, the APOGEE red giant star counts are corrected for their respective normalisation, via the reconstruction of the APOGEE DR16 selection function and using stellar evolution models, which enables the estimation of the stellar density profiles to good accuracy.

The paper is organised as follows: Section \ref{data} presents our selection of likely halo APOGEE red giant stars, and our criteria to determine N-rich stars residing in the stellar halo field. In Section \ref{method}, we briefly describe the density modelling procedure, based on the work by \citet{Bovy2016} and \citet{Mackereth2017}, including allowances for the APOGEE-2 selection function. In Section \ref{results}, we present the main results of the paper. Section \ref{discussion} includes the discussion of our results in the context of previous studies, for which we then summarize our results in Section ~\ref{conclusions}, and provide our conclusions.
\section{Data and Sample}
\label{data}
We use data from the sixteenth data release of SDSS-IV \citep{Masters2019}, which contains refined elemental abundances
\citep{Jonsson2020} for stars observed both by
the Apache Point Observatory Galactic Evolution Environment \citep[APOGEE][]{Majewski2017} and its successor, APOGEE-2, one of four Sloan Digital Sky
Survey--IV \citep[SDSS--IV,][]{Blanton2017} experiments. With twin spectrographs \citep{Wilson2019} mounted to the 2.5-m Sloan Telescope \citep{Gunn2006} at Apache Point Observatory in New Mexico and the 2.5-m du Pont Telescope at Las Campanas Observatory, APOGEE-2, includes high resolution (R$\sim$22,500), high signal-to-noise ratio (SNR>100 pixel$^{-1}$), near-infrared (1.5--1.7$\mu$m) spectra of over 450,000 stars sampling all parts of the Milky Way. These spectra are used to derive accurate stellar atmospheric parameters, radial velocities and the abundances for up to 25 atomic elements. Targets were selected
from the 2MASS point--source catalogue with a dereddened $(J
- Ks)$ $\geq$ 0.3 colour cut in up to three apparent $H$-band magnitude
bins. Reddening corrections were determined using
the Rayleigh--Jeans Colour Excess method \citep[RJCE;][]{Majewski2011}. Corrections are obtained by applying the method to the combined 2MASS
\citep{Skrutskie2006} and Spitzer-- IRAC surveys GLIMPSE--I,--II \citep{Churchwell2009}, and --3D when available, or 2MASS combined with WISE photometry \citep{Wright2010}. A more in--depth description on the
APOGEE survey, data reduction pipeline, and the target selection
can be found in \citet{Majewski2017}, \citet{Holtzman2015},
\citet{Nidever2015}, \citet{Jonsson2020}, \citet{Zasowski2013}, and \citet{Zasowski2017}. All
the APOGEE data products used in this paper are those output by the
standard data reduction and analysis pipeline. The data are first reduced with a custom pipeline (\citealp[][]{Nidever2015,Jonsson2020}). The data are then thoroughly checked, before being fed into the APOGEE
Stellar Parameters and Chemical Abundances Pipeline
(\citealp[ASPCAP;][]{Garcia2016,Jonsson2020}). ASPCAP makes use of a specifically
computed spectral library (\citealp[][]{Zamora2015,Holtzman2018,Jonsson2020}), calculated using a
customised \textit{H}-band line list \citep[][Smith et el, in prep]{Shetrone2015}, from
which then the outputs are then analysed, calibrated, and tabulated
\citep[][]{Holtzman2018}.

We make use of the distances for the APOGEE DR16 catalogue generated by \citet{Leung2019b}, using the \texttt{astroNN} python package \citep[for a full description, see][]{Leung2019}. These distances are determined using a previously trained astroNN neural-network, which predicts stellar luminosity from spectra using a training set comprised of stars with both APOGEE spectra and \textit{Gaia} DR2 parallax measurements \citep{Gaia2018}. The
model is able to predict simultaneously distances and account for
the parallax offset present in \textit{Gaia}-DR2, producing high precision,
accurate distance estimates for APOGEE stars, which match well
with external catalogues and standard candles.

The sample adopted in this study is restricted to stars contained in the statistical sample\footnote{Stars belonging to the statistical sample are those which were selected at random from 2MASS and are included in fields with completed cohorts.} of APOGEE DR16 located within fields with extinction information and that satisfy the following criteria: 0.5 < log \textit{g} < 3.5, 3500 < T$_{\mathrm{eff}}$ < 4750 K, --2.5 < [Fe/H] < --1, $d_{\mathrm{err}}$/$d$ < 0.2 (where $d$ and $d_{\mathrm{err}}$ are the distance and distance error, respectively). Here the T$_{\mathrm{eff}}$ and log \textit{g} cuts were made in order to target RGB stars whose C,N and Al abundances are unaffected by dredge up processes. We note here that we use the calibrated ASPCAP abundances, rather than those from other abundance pipelines (e.g., BACCHUS, astroNN).

We also remove stars that are in fields in close proximity to the Galactic Centre by excluding fields within 10$^{\circ}$ in $l$ from the Galactic Centre and |\textit{b}| < 10$^{\circ}$ for which dust extinction is most difficult to correct, as well as those which contained globular clusters (GCs) used for APOGEE calibration. Furthermore, since our aim is to study the Galaxy's stellar halo field population, we remove 1,781 stars in the APOGEE DR16 statistical sample that are known to reside in GCs, adopting the GC memberships established by \citet{Horta2020}. Lastly, since we are focused on selecting dissolved and/or evaporated SP GC stars in [N/Fe] space, we applied further cuts to ensure we were only considering stars where [N/Fe] is robustly determined, and removed any stars that had a \texttt{N\_FE\_FLAG} warning set by ASPCAP. We also ensured stars in our sample had reliable Fe, C and Al abundances by checking the \texttt{C\_FE\_FLAG} and \texttt{AL\_FE\_FLAG} warnings were set to zero, and by removing any stars with [Fe/H], [C/Fe], and/or [Al/Fe] set to --9999. Our final working halo sample is comprised of 1455 stars. 

\begin{figure*}
    \centering
    \includegraphics[width=\textwidth]{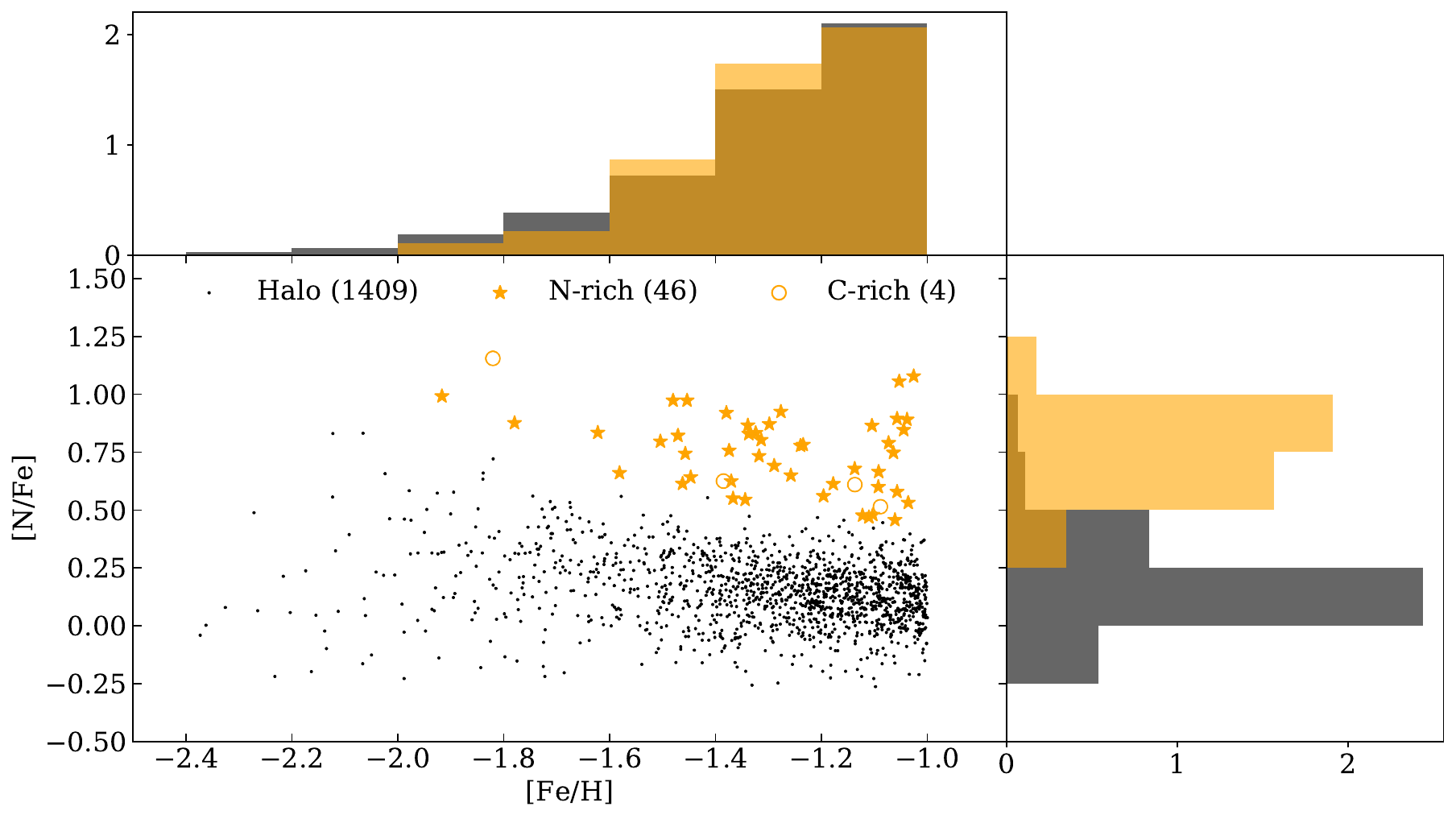}
    \caption{[N/Fe] vs [Fe/H] distribution of the halo field star sample (black dots) and the selected N-rich stars (orange stars) used in this work. Histograms highlight the MDF (top) and nitrogen distribution function (right) of both samples, normalised by the total star number of stars in each sample. Both samples share the same MDF, however can be clearly distinguished in the [N/Fe] distribution, with the mean N-rich star [N/Fe] value sitting approximately$\sim$0.7 dex higher than the mean halo field [N/Fe] abundance. Open circles at high [N/Fe] values are the N-rich stars that did not satisfy the [C/Fe] < 0.15 criterion. The numbers stated in brackets quantify the number of stars in each sample.}
    \label{fig_gmm}
\end{figure*}

Next, we define SP GC stars by identifying nitrogen rich (N-rich) stars present in the stellar halo field population of the Milky Way. To identify N-rich stars, we make use of the publicly available code \texttt{XDGMM}\footnote{https://github.com/tholoien/XDGMM} \citep{Holoien2017}, and fit a two component Gaussian mixture model (GMM) to our halo sample [N/Fe]--[Fe/H] distribution, with the expectation that N-rich stars will stand out in this plane. This software uses the extreme deconvolution (XD) algorithms \citep{Bovy2011} to identify components in an \textit{n} dimensional space, and allows us to determine the N-rich star field sample statistically by accounting for the uncertainties and correlations in the measurement errors. We then refine our N-rich star selection by only considering stars with carbon abundances below [C/Fe] < +0.15, to ensure that these stars present the low [C/Fe] abundances typical of SP GC stars, and remove four stars from the sample. Our final sample of N-rich stars is comprised of 46 stars. Fig.~\ref{fig_gmm} shows the distribution of the selected N-rich stars and the halo field population in the [N/Fe] - [Fe/H] plane, where the lowest N-rich star nitrogen abundance is [N/Fe]$\sim$+0.5, in agreement with values from other samples (\citealp[e.g.,][]{Martell2016,Schiavon2017,Nataf2019}). To check whether the metallicity distribution function (MDF; shown in the top panel of Fig. \ref{fig_gmm}) of our likely halo and N-rich star samples are statistically equal, we perform a two-sided Kolmogorov-Smirnov (KS) test, for which we obtain a resulting \emph{p-value} of approximately$\sim$0.75. Therefore, we can reject the null hypothesis that these two samples originate from different parent distributions and trust that they are statistically the same with a high degree of confidence; this indicates that the N-rich stars display a similar MDF as the halo field population. However, the N-rich star sample is clearly separated from the halo field sample in their nitrogen distribution function (right panel of Fig. \ref{fig_gmm}).

\begin{figure}
    \centering
    \includegraphics[width=\columnwidth]{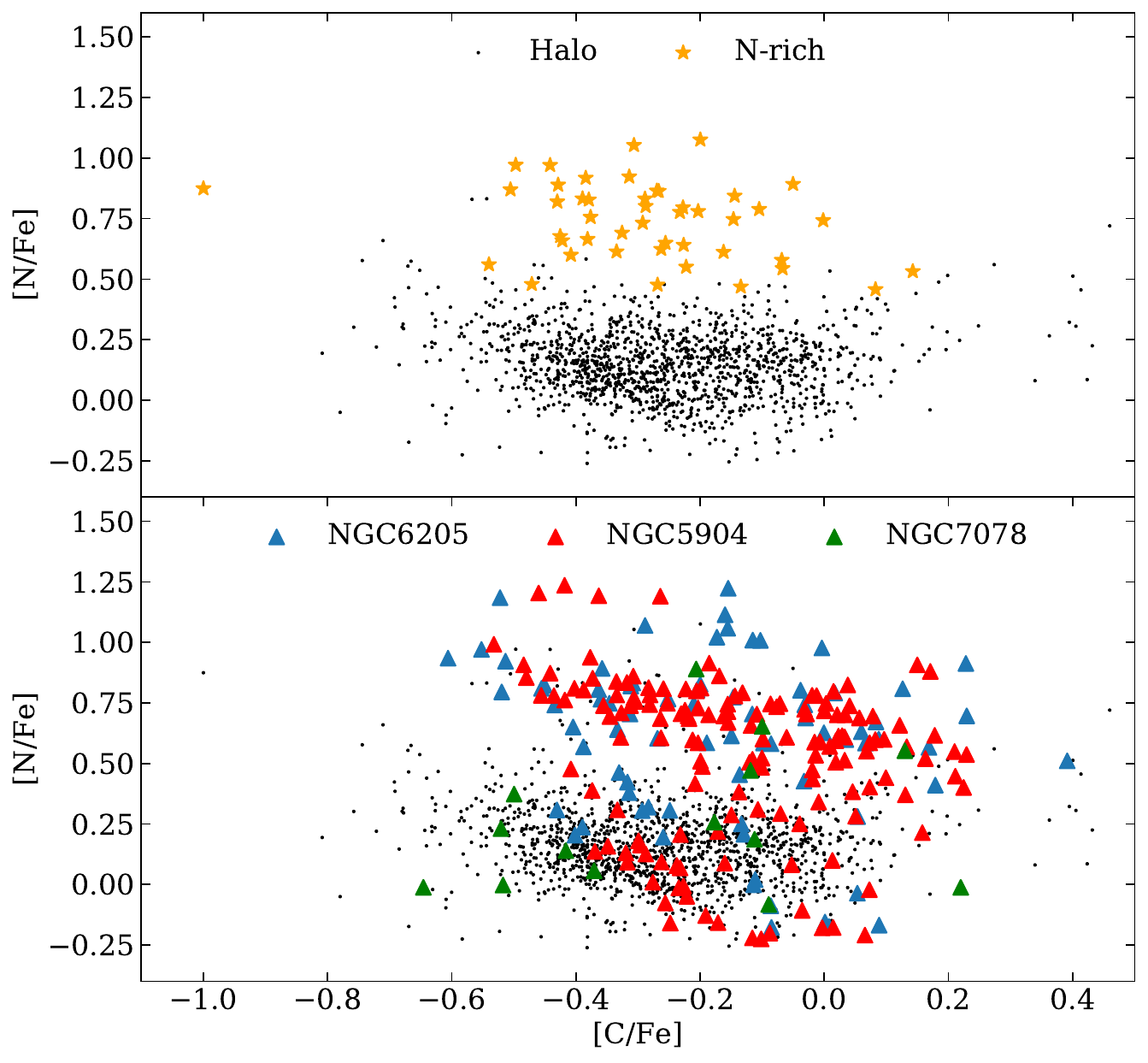}
    \caption{[N/Fe] vs. [C/Fe] distribution for the N-rich stars (top) and for stars from the APOGEE GC sample for three representative clusters (bottom) derived using the same procedure as in \citet{Horta2020}. The black points are the same in both panels, and represent the halo field population.  The N-rich star sample mimics the behaviour of SP GC stars of similar metallicity, occupying the same locus on this plane.  Note that the N-rich stars with the highest [C/Fe] have no counterparts in the FP sequence. This is due to a combination of factors.  Stars within that high [C/Fe] regime have higher T$_{\mathrm{eff}}$ and log$g$ and low [N/Fe], which makes CN lines weaker.  Moreover, there is a [N/Fe] floor in ASPCAP at --0.25.}
    \label{fig_cfes}
\end{figure}

\begin{figure}
    \centering
    \includegraphics[width=\columnwidth]{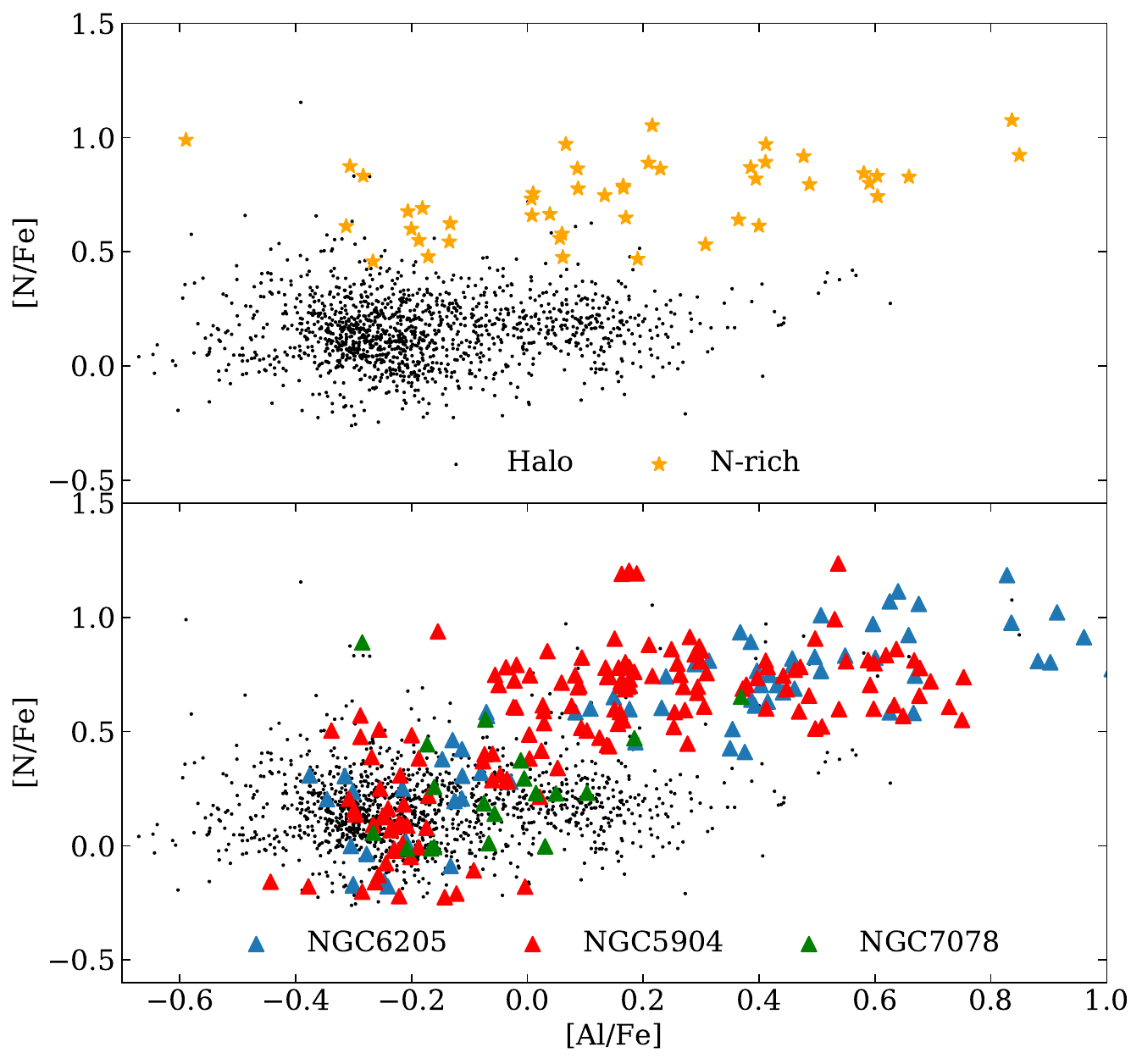}
    \caption{The same stars plotted from Fig.~\ref{fig_cfes} in the [N/Fe] vs. [Al/Fe] plane. Despite a small fraction of N-rich stars displaying [Al/Fe] < 0 dex, the majority occupy the same locus as the SP GC stars that display high [Al/Fe] abundances.}
    \label{fig_alfes}
\end{figure}

We plot the nitrogen abundances of the normal and N-rich halo field stars as a function of their respective carbon and aluminium abundances in Fig.~\ref{fig_cfes} and Fig.~\ref{fig_alfes}, respectively.  Those abundances are overlaid on data for stars residing in three GCs (NGC\,6205, NGC\,6904, NGC\,7078) whose metallicities span the same range as our N-rich star sample (i.e. --2.5 < [Fe/H] < --1), and for which we have substantial GC members, using the APOGEE GC sample from \citet{Horta2020}.  By comparing our N-rich star sample to GC stars, we aim to discern if our N-rich stars present the typical light element variations of SP GC stars (\citealp{Bastian2018}). The data for the GC members display two clear sequences in Figure \ref{fig_cfes}, corresponding to the loci of the FP and SP stars, with the latter typically showing [N/Fe] > +0.5.  Stellar evolution along the giant branch runs in parallel along each sequence, where more evolved, more luminous, lower $\log g$ stars have higher [N/Fe] and lower [C/Fe].  The well known N-C anti-correlation in GCs connects stars of same evolutionary stage in each sequence, so that FP stars have higher [C/Fe] and lower [N/Fe] than their SP counterparts.  With that in mind, the bottom panel suggests that the stars at the bottom of the SP sequence, with [N/Fe]$\sim$+0.5 and [C/Fe]$\sim$+0.2 have no counterparts in the FP sequence.  Those would be stars with lower [N/Fe] and higher [C/Fe] than the bottom of the FP sequence, which itself is located at [C/Fe]$\sim$+0.1 and [N/Fe]$\sim$--0.1.  We hypothesise that such stars are lacking in our sample due to a combination of effects.  Firstly, these stars relatively high T$_{\rm eff}$ and $\log g$, which makes CN lines, upon which nitrogen abundances rely, too weak for reliable N abundances.  Secondly, the ASPCAP grid has a [N/Fe] floor at --0.25.  

Figure \ref{fig_alfes} shows a correlation between [N/Fe] and [Al/Fe] for our N-rich star sample,
with the majority of these presenting [Al/Fe] > 0, again occupying the same locus as the SP GC stars. We also report that N-rich stars display a much greater mean [Al/Fe] ($\simeq$ 0.3) than the normal halo field population ([Al/Fe] $\simeq$ --0.2). We also checked whether the N-rich population differed from its normal halo counterpart in terms of its neutron-capture element abundances, since those were found to be enhanced in SG GC stars by \citet{Cunha2017}.  The only neutron-capture element for which APOGEE DR16 provides abundances for our sample is Ce. We found no statistically significant difference in [Ce/Fe] between the two samples.

In a recent paper, \citet{Trincado2019} suggest that SP GC stars must have [Al/Fe]>+0.5 in order to be considered true SP GC star candidates. In fact, adoption of such a restrictive criterion would lead to a severe underestimate of the N-rich population in both our field and GC sample. Stars belonging to NGC\,6205, NGC\,5904, and NGC\,7078 from the \citet{Horta2020} GC sample display a clear bimodality in the carbon-nitrogen plane, where the SP GC stars occupy a locus positioned at higher [N/Fe], congregating above their FP counterparts. This bimodality is also observed in the aluminium-nitrogen plane. In the bottom panel of Fig. \ref{fig_alfes}, the SP GC stars (i.e. the filled triangles not overlapping with the halo population) clearly present [Al/Fe] abundances that fall below [Al/Fe]=+0.5. Thus, since our N-rich star sample occupies the same locus as the SP GC stars in both the carbon-nitrogen and aluminium-nitrogen planes, applying an [Al/Fe] > +0.5 cut to our sample would lead to a similarly severe underestimate in the number of N-rich stars in the field. In addition, it has been shown by several authors \citealp[e.g.,][]{Carreta2009,Nataf2019,Meszaros2020} that the extent of the Mg-Al anti-correlation varies from GC to GC, making a strict cutoff in Al difficult to implement.


\begin{figure*}
    \centering
    \includegraphics[width=\textwidth]{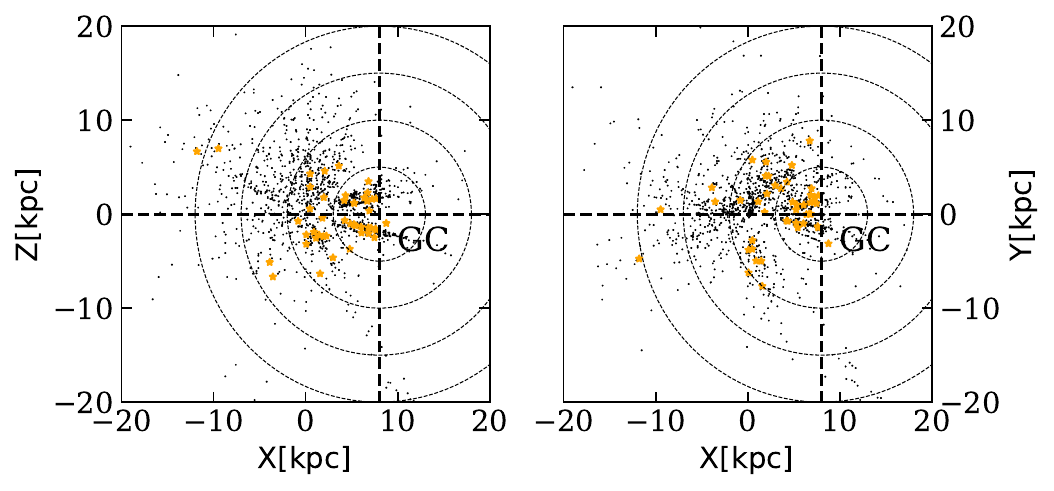}
    \caption{Spatial distribution in heliocentric \textit{X} - \textit{Z} and \textit{X} - \textit{Y} of our APOGEE DR16 field (black dots) and N-rich star (orange stars) samples between --2.5 < [Fe/H] < --1. The crosshair signifies the position of the Galactic Centre. The dashed circled lines denote spherical radius bins of 5 kpc in size, signifying a 5, 10, 15, and 20 kpc distance from the Galactic Centre.}
    \label{fig_spatial}
\end{figure*}

Finally, we show the spatial distribution of the halo field population and the N-rich star sample in Fig.~\ref{fig_spatial}, represented with the same symbols as in Fig.~\ref{fig_gmm}. Before correcting for the APOGEE selection function or modelling the stellar density, it is immediately clear that the halo field population (black) occupies a vast range of heights above the Galactic midplane, reaching values of \textit{Z}$\sim$20 kpc. The same applies to the N-rich stars, however these do not display as large a Galactocentric distance range. Interestingly, we find that a significant fraction of the N-rich stars occupy a position close to the Galactic Centre, within an approximate radius of $R_{\mathrm{GC}}$$\sim$3 kpc. The remaining N-rich stars are scattered at higher \textit{Z} and \textit{Y} values. It has already been shown that N-rich stars make up a significant fraction of the inner Galaxy \citep{Schiavon2017} and to a lesser extent of the outer halo (\citealp{Martell2016,Koch2019}), therefore our initial findings are in agreement with results from previous studies.

\section{Density Modelling}
\label{method}
In this section, we briefly describe the method employed for fitting the underlying number density of stars in the Milky Way's halo from APOGEE observations, which we represent here as $\nu_{*}$($X$,$Y$,$Z$|$\tau$), in units of stars kpc$^{-3}$. The computation of this quantity requires allowances to be made for the survey selection function, which is non-trivial due to the pencil beam nature of APOGEE, the presence of inhomogeneous dust extinction along the lines of sight, the target selection criteria imposing different \textit{H} magnitude limits, and the use of tracers, RGB stars, that are not standard candles. Our main goal is to determine density laws that describe the spatial distributions of the stellar halo and the N-rich samples separately.  In this way, the number of observed normal field and N-rich stars can be converted into actual densities, in units of stars pc$^{-3}$ at the solar radius ($N_{R_0}$). This density law can then be used to estimate the number of halo or N-rich stars APOGEE would have observed if it covered the full sky. Moreover, using stellar evolution models we can estimate the mass density. This value can then be estimated within a chosen volume by integrating an accurate density law for the halo and N-rich star samples, respectively. It is also straightforward to use the separate density laws to determine the ratio of mass density between N-rich stars and halo field stars, resulting in the fractional number of N-rich to total halo field stars, given as a percentage, as a function of position in the Galaxy. By following this procedure, we aim to determine if the ratio of N-rich stars is constant throughout the Milky Way, and if N-rich stars follow a similar density profile to that of the stellar halo field population.

Our methodology is an adaptation of that used by \citet{Mackereth2017} and \citet{Bovy2012a,Bovy2014,Bovy2016}. We employ a modified version of their publicly available code\footnote[3]{Available at https://github.com/jmackereth/halo-mass and https://github.com/jobovy/apogee-maps, respectively.}. Despite our method following that of these previous studies, we summarize the key steps here for clarity and completeness. For a full description of the fitting method, we refer the interested reader to Section 3.1 from \citet{Mackereth2019b}. The main steps are summarised as follows:

(1) We fit different density functional forms for the density profile to the APOGEE N-rich star sample using a maximum likelihood fitting procedure, based on the assumption that star counts are well modelled by an inhomogeneous Poisson point process\footnote[4]{ See \citet{Daley2003} for a more detailed definition of an inhomogeneous Poisson point process.}, which takes into account the APOGEE selection function. For the halo sample, we adopt the functional form from \citet{Mackereth2019b} (see Eq. \ref{eq1} in Section \ref{halo_models}) for two reasons: Firstly, the sample employed in their study is extremely similar to that used here; and secondly, their model describes our data well (see Section \ref{results}).

(2) We determine the best-fitting density model for our N-rich star sample by calculating the Bayesian Information Criterion \citep[BIC][]{Schwarz1978} and the logarithmic maximum likelihood value, ln($\mathcal{L}_{\mathrm{max}}$), for each model, and choosing that with the lowest BIC value (see Section \ref{n-rich_models} and Section \ref{bic} for a description of the models tested and the selection of our best fitting model, respectively). Our selected halo density profile includes a single exponential disc profile to account for the contribution of the (thick) disc component to our sample. We find that by including this disc fraction parameter to our model, we are able to quantify the fraction of the sample that belongs to the (thick) disc and that would affect the fitting procedure. Moreover, given our metallicity cut, the contribution by the thin disk to our sample is negligible, so we do not include that component in our model. Using this density law for the halo, and the resulting best fitting model for the N-rich stars, we initiate a Markov-Chain Monte-Carlo (MCMC) sampling of the posterior probability distribution function (PDF) of the parameters fitted in each density law. We adopt the median and standard deviation of one-dimensional projections of the MCMC chain as our best-fit parameter values and uncertainties.

(3) As the fitting procedure does not fit for the normalisation of the density \textit{N$_{\mathrm{R_{0}}}$} (the number surface density of halo or N-rich stars at the solar neighbourhood in stars pc$^{-3}$), we compute this quantity by comparing the observed number of halo or N-rich stars in the metallicity bin adopted to that which would be observed in APOGEE for the fitted density model if \textit{N$_{\mathrm{R_{0}}}$} = 1 star pc$^{-3}$. We then employ this number-density quantity (\textit{N$_{\mathrm{R_{0}}}$}) to obtain the true number of stars given by our best-fitting halo or N-rich density model within a given volume. This number density can then be converted into a mass density using stellar evolution models. Since our aim is to determine the contribution of N-rich stars as a function of Galactocentric distance, we take the ratio of the mass densities yielded for the N-rich star sample and the halo sample. This allows us to 
compare the ratio of GC dissolution in different spatial regions of the Galaxy, and potentially place constraints on the origin of the contribution of dissolved GC stars to the Galaxy's stellar halo mass budget, and the mass assembly history of the Milky Way.

Our method corrects for the effects induced by interstellar extinction using combined 3D maps of the Milky Way derived by \citet{Marshall2006} for the inner disc plane and those derived for the majority of the APOGEE footprint by \citet{Green2019}, adopting conversions \textit{A$_{\mathrm{H}}$}/\textit{A$_{\mathrm{K_{s}}}$} = 1.48 and \textit{A$_{\mathrm{H}}$}/\textit{E}(\textit{B - V}) = 0.46 (\citealp{Schafly2011,Yuan2013}). Any fields with no dust data (10 fields) were removed from the analysis. We adopt the combination of \citet{Marshall2006} and \citet{Green2019} dust maps because \citet{Bovy2016}, who assessed the relative merits and limitations of a number of available maps in the literature, determined that this combination of dust maps provides the best fits when performing density modelling on APOGEE data for a trace population.

\subsection{Halo Density Model}
\label{halo_models}
We aim to discern the profile that best describes our halo field sample. Recent work modelling the Galactic stellar halo using RR Lyrae stars found that the stellar halo is well modelled by a triaxial density ellipsoid, angled at roughly 70$^{\circ}$ with respect to the axis connecting the Sun and the Galactic Centre \citep{Iorio2018,Iorio2019}, but can also be well modelled by a single power law (SPL). Along the same lines, \citet{Mackereth2019b} used the APOGEE DR14 data and the corresponding APOGEE selection function, and found that the mono-abundance populations in the [Fe/H] range selected in that work are well modelled by a triaxial SPL density ellipsoid with a "cut-off" term. The cut-off term accounts for our ignorance regarding the extent of the sample outside of the observed range, and includes a scale parameter $\beta$ corresponding to the scale length (such that \textit{h$_{\mathrm{r_{e}}}$} = 1/$\beta$). This density profile also includes a disc term, modelled by an exponential disc profile with scale height \textit{h$_{\mathrm{z}}$} = 0.8 kpc and scale length \textit{h$_{\mathrm{R}}$} = 2.3 kpc \citep{Mackereth2017}, to account for any contamination from the thicker components of the high [$\alpha$/Fe] disc, parameterised by the factor \textit{f$_{\mathrm{disc}}$}. Hereafter, we refer to this model as TRI-CUT-DISC.

Since we are modelling a similar metal-poor halo sample as that one modelled in \citet{Mackereth2019b}, with a very similar MDF, we adopt the functional form from their best-fitting profile to model the stellar halo field population. That density profile is given by:
\begin{equation}\label{eq1}
 v_{*}(r_{\mathrm{e}}) \propto (1 - f_{\mathrm{disc}})  r_{\mathrm{e}}^{-\alpha}  \exp(-\beta r_{\mathrm{e}}) + f_{\mathrm{disc}}v_{*,\mathrm{disc}}.
\end{equation}
The density is normalised such that the density at the Sun (i.e., $ v_{*}(r_{\mathrm{e},0})$) is equal to unity, and $r_{\mathrm{e}}$ defines the ellipsoidal surfaces on which the density is constant, given by:
\begin{equation}\label{eq2}
    r_{\mathrm{e}}^{2} = X^{2}_{\mathrm{g}} - \frac{Y_{\mathrm{g}}^{2}}{p^{2}} - \frac{Z_{\mathrm{g}}^{2}}{q^{2}},
\end{equation}
where \textit{p} and \textit{q} describe the \textit{Y$_{\mathrm{g}}$}-to-\textit{X$_{\mathrm{g}}$} and \textit{Z$_{\mathrm{g}}$}-to-\textit{X$_{\mathrm{g}}$} axis ratios, respectively, and (\textit{X$_{\mathrm{g}}$},\textit{Y$_{\mathrm{g}}$},\textit{Z$_{\mathrm{g}}$}) are the Cartesian coordinates relative to the Galactic Centre. The best-fitting model found in \citet{Iorio2018} included an allowance for the variation in \textit{q} with $r_{\mathrm{e}}$, with a scale length of$\sim$13 kpc. In this work we assume \textit{q} does not vary with elliptical radius, and constrain both shape parameters \textit{q} and \textit{p} to be < 1, forcing the longest axis to be that in the \textit{X$_{\mathrm{g}}$} direction. As in \citet{Mackereth2019b}, we allow the orientation of the density ellipsoid to vary, applying a transformation defined by the unit vector \textit{\^{z}} along the transformed \textit{Z} axis, and the angle of rotation (from the original \textit{X} axis) of the ellipsoid about this transformed axis, $\phi$. In practice, \textit{\^{z}} is sampled uniformly (with the MCMC algorithm) by de-projecting samples from an equal-area rectangular projection of the unit sphere. Employing this method, \textit{\^{z}} is represented by two parameters, $\eta$ and $\theta$, such that \textit{\^{z}} = ($\sqrt{1 - \eta^{2}}$cos$\theta$, $\sqrt{1 - \eta^{2}}$sin$\theta$, $\eta$). Here $\eta$ is sampled uniformly between --1 and 1, and $\theta$ between 0 and 2$\pi$. Such a transformation generally has little impact on the measurement of the total mass, as the parameters tend to define an ellipsoid which is roughly aligned with the Sun's position. For all parameters we adopt uninformative priors, and set the allowed exponent $\alpha$ and parameter $\beta$ range to be positive. For the remaining parameters, only a range of [0,1] is permitted.

\subsection{N-rich star density models}
\label{n-rich_models}
With no pre-conception of the density distribution of N-rich stars in the Galaxy, it is important to test several different density models for the N-rich sample, and statistically evaluate which model best fits the data. In this subsection, we describe the range of density models fitted to the N-rich star data, and provide the analytical function for clarity and completeness. 

We begin by fitting the simplest model to the N-rich star sample, a spherical power law (SPL). This model is described analytically as:
\begin{equation}\label{eq3}
    \nu_{*}(r_{\mathrm{e}}) \propto r^{-\alpha},
\end{equation}
where $r$ is given by Eq. \ref{eq2} when equating $p$ and $q$ to unity. This profile assumes the density is prescribed by spherical shells, with no flattening along any axis. A flattening parameter along the \emph{Z} axis ($q$) can be introduced to obtain an axisymmetric profile (AXI), which is described by Eq. \ref{eq3}, assuming $r$ is given by Eq. \ref{eq2}, and equating $p$ to one. Similarly, a separate flattening parameter along the \emph{Y} axis ($p$) can be introduced to obtain a triaxial profile (TRI), governed by the shape of a triaxial ellipsoid. Alongside these three single power law density models, we choose also to fit the model described in Section \ref{halo_models} (TRI-CUT-DISC), employed to fit the stellar halo field data.

In addition to the profiles mentioned above, we fit two further profiles that are not part of the single power law family. The first is an exponential disc profile from \citet{Mackereth2017} (DISC), given by:
\begin{equation}\label{eq4}
    \nu_{*}(r_{\mathrm{e}}) \propto \exp(-h_{\mathrm{R}}(R-R_{0}) - h_{\mathrm{z}}|z|),
\end{equation}
where $R$ and $z$ are the radial and vertical axes in a cylindrical coordinate system, respectively, $h_{\mathrm{R}}$ and $h_{\mathrm{z}}$ are the radial and vertical scale lengths, and $R_{0}$ is the Galactocentric radius of the Sun.  We believe that by testing a disc profile, we will be able to discern if the N-rich star sample is better described by a disc density model, and thus if it has a high contribution to the Galactic disc component.
The second additional profile fitted is a broken power law (BPL). This density model resembles that of a single power law (see Eq \ref{eq3}), however has a break radius denoting a change in the slope of the model, given by a change in the value of $\alpha$. The broken power law is given by:
\begin{equation}\label{eq5}
    \nu_{*}(r_{\mathrm{e}}) \propto \Bigg\{ 
    \begin{matrix}
    r^{-\alpha_{\mathrm{in}}} & \mathrm{if}& R < R_{\mathrm{break}}\\
    r^{-\alpha_{\mathrm{out}}} & \mathrm{if}& R > R_{\mathrm{break}},
    \end{matrix}
\end{equation}
where $\alpha_{\mathrm{in}}$ is the power law exponent inside the break radius (i.e. $R_{\mathrm{break}}$), and $\alpha_{\mathrm{out}}$ is the exponent outside the break radius. By fitting a broken profile, we will be able to test the hypothesis in which the N-rich star sample is governed by a break between an inner and outer component, potentially linked to the different N-rich star ratios found in previous studies for different spatial regions.

For all the models presented we adopted uninformative priors, as performed in Section \ref{halo_models}. For the parameters of the exponential disc profile, we set a permitted scale length range of [0,10] (kpc). For the permitted range of the break radius of the broken power law, we permit values between [0,20] (kpc), a range of Galactocentric distances that is relatively well covered by our data.

\subsection{Identification of best-fitting density model}
\label{bic}
In this subsection, we describe the methodology employed to select the best fitting N-rich star profile from the density models described in Section \ref{n-rich_models}. We choose the Bayesian Evidence ratio as our figure of merit to identify the best-fitting model. We assume that the posterior distributions are nearly Gaussian, and that therefore the Bayesian Evidence ratio can be approximated by the Bayesian Information Criterion \citep[BIC;][]{Schwarz1978}, defined as:
\begin{equation}\label{eq6}
    BIC = d(\tau)ln(N_{*}) - 2ln (\mathcal{L}_{\mathrm{max}}),
\end{equation}
where \textit{d}($\tau$) is the number of free parameters in the density model, \textit{N}$_{*}$ is the number of stars in our sample, and ln($\mathcal{L}_{\mathrm{max}}$) is the logarithm of the maximum value of the likelihood.

The best fitting model is that for which the BIC value is the lowest. The BIC statistic penalises models with a large number of free parameters, such that for two models with the same ln($\mathcal{L}_{\mathrm{max}}$) value, the one fewer free parameters is preferred.
The results from this comparison are listed in Section \ref{results_nrich}, displayed in Table \ref{tab1}. For the fit of each profile to the N-rich star data, we refer the reader to Appendix \ref{fitting_Nrich}.

\subsection{Mass estimation}
We follow the work by \citet{Bovy2012b} and \citet{Mackereth2019b}, and estimate the mass for our halo and N-rich samples. Although the methodology is fully described in \citet{Mackereth2019b}, we briefly explain the procedure for estimating the mass within an APOGEE star sample for clarity. Upon finding a best-fitting model and its associated uncertainty for a distribution of stars, the measurement of the mass can be computed by employing the normalisation of the rate function:
\begin{multline} \label{eq7}
    \lambda(O|\tau) = \nu_{*}(X,Y,Z|\tau) \times  |J(X,Y,Z;l,b,D)| \\
\times  \rho(H,[J - K_{s}]_{0},\mathrm{[Fe/H]}|X,Y,Z) \times S(l,b,H),
\end{multline}
where $\nu_{*}(X,Y,Z|\tau)$ is the stellar number density in rectangular coordinates, in units of stars kpc$^{-3}$. \textit{|J(X,Y,Z;l,b,D)|} is the Jacobian of the transformation from rectangular (\textit{X,Y,Z}) to Galactic (\textit{l,b,D}) coordinates, and $\rho(H,[J - K_{s}]_{0},\mathrm{[Fe/H]}|X,Y,Z)$ represents the density of stars in magnitude, colour, and metallicity space given a spatial position (\textit{X,Y,Z}), in units of stars per arbitrary volume in magnitude, colour, and metallicity space. \textit{S}(\textit{l,b,H}) is the survey selection function (\citealp[see][for details]{Bovy2016,Mackereth2017}), which denotes the fraction of stars successfully observed in the survey's colour and magnitude range, and includes dust extinction effects.

We perform this mass-estimation procedure on both our total halo and N-rich star samples by calculating the number of stars seen by APOGEE for a given density model normalised to unity at the Solar position, \textit{N}(\textit{$\nu_{*,0}$} = 1). That number is obtained by integrating the rate function over the observable volume of the survey. This integral is given by:
\begin{multline}\label{eq8}
        N(\nu_{*,0} = 1) = \int_{\mathrm{fields}} d \textup{field}\, dD\, \lambda(\mathrm{field},D) = \\
        \int d \textup{field}\, d\mu\,\frac{D^{3}(\mu)\mathrm{log}(10)}{5}\nu_{*}([R,\phi,z](\mathrm{field},\mu|\tau)) \times \mathfrak{S}(\mathrm{field},\mu),
\end{multline}
where the density and effective selection function (namely, $\mathfrak{S}$(field,$\mu$)) are calculated along APOGEE sightlines on a grid linearly spaced in distance modulus $\mu$. Since the
true number of observed stars is given by:
\begin{equation}
{\rm N}_{\mathrm{obs}} \,=\, {\rm A\, N}(\mu_{*,0}=1),
\end{equation}
comparison of the expected number count for a normalised density model with the
true observed number of stars in the sample provides the proper
amplitude, A, which is then equivalent to the true number density of
RGB stars at the Sun, $\mu_{*,0}$, when considering either our halo or N-rich star sample.

The number density in concentric triaxial ellipsoids can also be calculated as a function of Galactocentric distance. Once a best-fitting model for a specific sample of APOGEE stars is estimated, the number density can be computed on a grid of $\nu_{*}$([R,$\phi$,$\theta$]) over a chosen volume. We can then compute this value as a function of Galactocentric radius by integrating along the $\phi$ and $\theta$ axis, which can later be converted into stellar-mass density using stellar evolution models. In either case, the number counts in RGB stars can be converted into the mass of the entire underlying population. To do so, we use the PARSEC isochrones (\citealp{Bressan2012,Marigo2017}), weighted with a \citet{Kroupa2001} IMF. The average mass of RGB stars $\langle$$M_{\mathrm{RGB}}$$\rangle$ observed in APOGEE is then calculated by applying the same cuts in (\textit{J} -- $K_{s}$)$_{0}$ and $\log{g}$ to the isochrones. The fraction of the stellar mass in giants, $\omega$, is given by the ratio between the IMF weighted sum of isochrone points within the RGB cuts and those of the remaining population. The conversion factor between RGB number counts and total stellar mass can then simply be calculated using:
\begin{equation}\label{eq9}
    \chi(\mathrm{[Fe/H]}) = \frac{\langle M_{\mathrm{RGB}} \rangle (\mathrm{[Fe/H]})}{\omega(\mathrm{[Fe/H]})}.
\end{equation}
As explained in \citet{Mackereth2019b}, the factor for each field and each selection in [Fe/H] is computed by adjusting the limit in ($J$ -- $K_{s}$)$_{0}$ to reflect the minimum ($J$ -- $K_{s}$)$_{0}$ of the bluest bin adopted in that field, and only considering isochrones that fall within --2.5 < [Fe/H] < --1. The edges in colour binning for each field are accounted for by the integration over $\rho$[($J$ -- $K_{s}$)$_{0}$,$H$] for the effective selection function. \citet{Mackereth2019b} used Hubble Space Telescope photometry to show that the factors determined using their method are reliable against systematic uncertainty arising from the choice of stellar evolution models. 

Once the normalisation for our sample is obtained, we integrate the normalised density models described by 400 samples from the posterior distributions of their parameters to attain the total mass within a population, and the total mass as a function of Galactocentric radius for that same population. To avoid over-extrapolation from our fits to the halo density, we only integrate for the mass within a 1.5 < $r$ < 15 kpc range, for which our data are well contained.

\section{Results}
\label{results}
\subsection{Fit to the halo sample}
\label{halofit}
\begin{figure*}
    \centering
    \includegraphics[width=\textwidth]{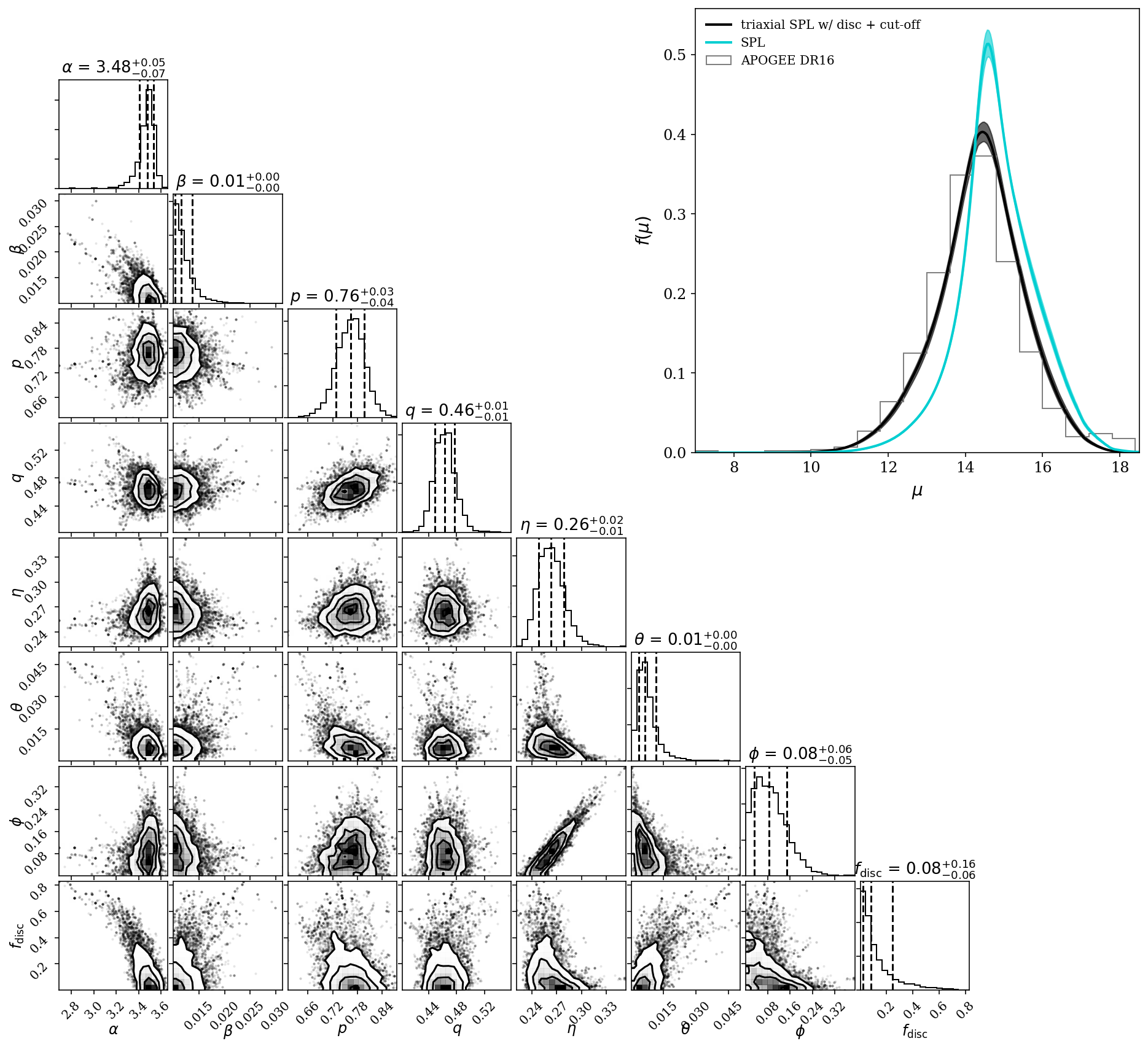}
    \caption{Corner plot showing the posterior 10,000 samples of the parameters for the adopted triaxial disc cut-off single power law (TRI-CUT-DISC) model when fit to the full statistical likely halo sample (1409 stars) between --2.5 < [Fe/H] < --1. The posterior distributions are well behaved, and converge to a median value. The best fit profile has a mildly steep power law of $\alpha$ $\simeq$ 3.5, and is slightly flattened along its \textit{Y} and \textit{Z} axes. The inset panel (top right) shows the distance modulus $\mu$ distribution predicted by the best fit model (black), and the best fit SPL (blue) of similar $\alpha$ $\simeq$ 3.2, where the thickness of each line signifies the 1-$\sigma$ spread. The grey histogram represents the real APOGEE data, and shows that the TRI-CUT-DISC profile provides a better quality fit.}
    \label{fig_cornerhalo}
\end{figure*}

We perform the fitting procedure described in Section \ref{method} to our APOGEE halo sample, to ensure that the prescribed density provides a good fit to the data. Since it has recently been shown that the APOGEE DR14 star sample between --2.5 < [Fe/H] < --1 is well defined by a triaxial SPL density ellipsoid, which included a cut-off term and freedom to rotate around its axis \citep{Mackereth2019b}, we choose to fit this profile to our sample. This profile also included a parameter to account for the fraction of the disc in our sample, described by an exponential disc profile. The resulting best-fitting profile, obtained running 10,000 realisations of the model, follows a moderately steep power law, yielding a value of $\alpha$ $\simeq$ 3.48 $^{+0.05}_{-0.07}$, with a slight flattening along the \textit{Y} axis ($p$ = 0.76 $^{+0.03}_{-0.04}$), and slightly more flattened along the \textit{Z} axis, with $q = 0.46 \pm 0.01$, which is in good agreement with previous results (\citealp[e.g.,][]{Deason2011,Xue2015,Iorio2018,Mackereth2019b}). We find the $\beta$ cut-off parameter to be$\sim$0.01, indicating a scale length which is well outside the range of our data, of the order of $h_{\mathrm{r_{e}}}$$\sim$100 kpc, and that the triaxial ellipsoid is consistent with a minor rotation around the \emph{Z} axis, $\theta$ = 3.6 $\pm$ 0.5 degrees, and a slighly larger rotation along the \emph{X} axis, $\phi$ = 14 $^{+11}_{-9}$  degrees. Our results also show that our halo sample has low contamination from the disc, given by the low $f_{\mathrm{disc}}$ = 0.08 $^{+0.16}_{-0.06}$ value. The samples from the posterior distribution of parameters given by the data are illustrated in a corner plot in Fig.~\ref{fig_cornerhalo}, which shows that the posterior distributions are well behaved and converge towards a single value. Alongside the samples, we show in the top right panel the observed distance modulus $\mu$ distribution as predicted by a triaxial disc cut-off profile (black) and a spherical power law (blue) with an exponent that best fits the data ($\alpha$ $\simeq$ 3.2), compared to the real APOGEE data. Given that our triaxial disc cut-off power law fits our data well, and yields parameter estimates within agreement with values from previous studies, we choose to use this model as our best fit density profile.

It is interesting to compare our best estimate of the halo enclosed mass and normalisation with those by \citet{Mackereth2019b}. Using the triaxial disc cut-off power law profile and the APOGEE star counts, we obtain a halo normalisation $\rho_{*,0}$ = 1.3 $^{+0.1}_{-0.2}$ $\times$ 10$^{-4}$ M$_{\odot}$ pc$^{-3}$, and a total stellar halo mass M$_{*,\mathrm{halo}}$ = 8.3 $^{+1.5}_{-1.3}$ $\times$ 10$^{8}$ M$_{\odot}$, by using Eq ~\ref{eq8} and integrating over the full observable volume (i.e. r$\sim$1.5-15 kpc). We find our total stellar halo mass estimate to be larger by a factor of 1.5 than that obtained by \citet{Mackereth2019b} for the same density profile using APOGEE DR14 data, and suggest it is likely due to the different volumes employed to integrate the density (\citet{Mackereth2019b} integrated the halo volume within a Galactocentric distance ranging from 2 < r < 70 kpc). Furthermore, within the uncertainties, we find our total halo mass estimate to be in borderline agreement with the estimated 4-7 $\times$ 10$^{8}$ M$\odot$ from the review by \citet{Bland-Hawthorn2016}, and to be smaller than the$\sim$10$^{9}$ M$\odot$ estimate presented in the work of \citet{Deason2019}, or the sum of individual MAPS in \citet{Mackereth2019b}.

\subsection{Fitting the N-rich star sample}
\label{results_nrich}
In this subsection, we describe the density modelling of the chemically defined N-rich stars. We test six different density profiles ranging in the number of free parameters, and calculate the ln($\mathcal{L}_{\mathrm{max}}$) and BIC values in order to estimate the best fit density profile. The models of choice are described in detail in Section \ref{n-rich_models}, and are summarised as follows: a single power law (SPL), an axisymmetric power law (AXI), a triaxial power law (TRI), the triaxial disc cut-off profile used to fit the halo sample (TRI-CUT-DISC), a broken power law (BPL), and an exponential disc profile (DISC). 
Employing the method described in Section \ref{bic}, we have obtained the BIC and ln($\mathcal{L}_{\mathrm{max}}$) for the six models tested, which are listed in Table \ref{tab1}. The fit of each profile to the N-rich star data is shown in Fig. \ref{fig_nrich_fits} in Appendix \ref{fitting_Nrich}. We draw 10,000 samples from the posterior of each model, for which we take the median value as the best fit parameters for each profile. We then compute the BIC and ln($\mathcal{L}_{\mathrm{max}}$). As is evident from the resulting BIC values from Table \ref{tab1} (and from Fig. \ref{fig_nrich_fits}), the AXI profile is the best fitting profile. However, we find the N-rich star sample can also be well fitted by the TRI-CUT-DISC and TRI profiles.

\setlength{\tabcolsep}{18pt}
\begin{table}
\begin{tabular}{ |p{2cm}|p{1.cm}|p{1.5cm}|  }
\hline
Density profile & $\Delta$ BIC & $\Delta$ ln($\mathcal{L}_{\mathrm{max}}$)\\
\hline
\hline
  SPL &34.3 & 19.3\\
  \textbf{AXI} &\textbf{0.0}& \textbf{0.0}\\
  TRI& 2.2& 0.5\\
  TRI-CUT-DISC& 10.0& 2.6\\  
  BPL& 60.9& 17.6 \\
  DISC& 9.1& 12.0 \\
  \hline
\hline
\end{tabular}
\caption{Summary of the results for a sample of density profiles used to fit the N-rich star sample. The models tested are: a single power law (SPL), an axisymmetric power law (AXI), a triaxial power law (TRI), a triaxial power law with a disc and cut-off term (TRI-CUT-DISC), a broken power law (BPL), and an exponential disc (DISC). For each model, we report the Bayesian Information Criterion (BIC) and maximum logarithmic likelihood differences between the best fit model (bold) and the remaining models, calculated by taking 10,000 MCMC samples and using the median posterior parameter values. }\label{tab1}
\end{table}

Once we obtain our best fitting profile, we perform the density fitting procedure described in Section \ref{method} on the N-rich star sample, and obtain a density profile that provides an exquisite fit to the data. After running 10,000 MCMC iterations, we obtain a density profile with a slope of $\alpha = 4.47^{+0.23}_{-0.22}$, which is much steeper than that of the halo profile. Furthermore, we find the N-rich profile to have roughly the same flattening along the $Z$ axis as the halo profile (namely, \textit{q} = 0.47$^{+0.05}_{-0.04}$).

\begin{figure*}
    \centering
    \includegraphics[width=\textwidth]{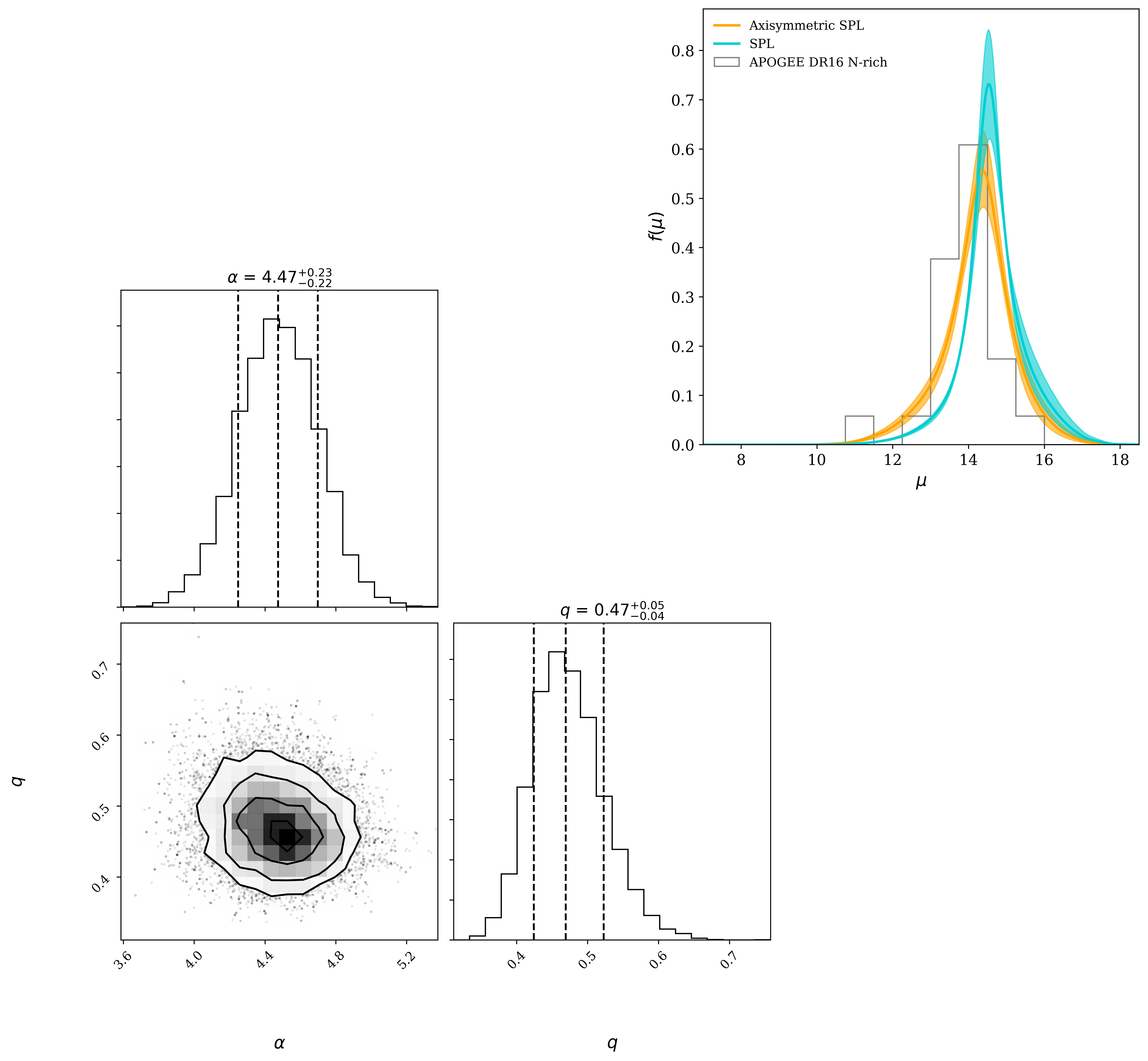}
    \caption{The same as in Fig. \ref{fig_cornerhalo}, however for the 46 N-rich stars displayed in Fig. \ref{fig_gmm} fitted by an axisymmetric single power law (AXI). As in Fig. \ref{fig_cornerhalo}, the posterior distributions are well behaved. The best fit model displays a steeper exponent $\alpha$ $\simeq$ 4.5 in comparison to the halo sample, but shows that it is similarly flattened along the \textit{Z} axis. As in Fig. \ref{fig_cornerhalo}, the inset panel displays the predicted distance modulus $\mu$ distribution predicted by the best fit profile (orange) and a best fit SPL with $\alpha$ $\simeq$ 3.9 (blue), compared to the N-rich star data.}
    \label{fig_cornernrich}
\end{figure*}

The resulting samples from the posterior distribution of parameters given by the data are shown in the corner plot illustrated in Fig. \ref{fig_cornernrich}, which shows that the posterior distributions are well behaved and converge to a median value. In addition to the resulting MCMC samples, we show the distance modulus $\mu$ distribution predicted by the best fit model (orange) and a spherical power law (blue) with an exponent that best fits the data ($\alpha$ $\simeq$ 3.9), compared to the real N-rich star APOGEE data. Using the best profile and APOGEE N-rich star counts, we obtain an N-rich normalisation $\rho_{0,\mathrm{N-rich}}$ = 3.1 $^{+0.6}_{-0.5}$ $\times$ 10$^{-6}$ M$\odot$ pc$^{-3}$. Given the reasonable fit to the N-rich star data by the axisymmetric power law, we choose to use this profile as our best fitting N-rich density profile. However, we point out that the TRI and TRI-CUT-DISC profiles also provide a good quality fit to the N-rich star data, and yield a normalisation value that is consistent, within the uncertainties, with the value obtained for the AXI profile. Moreover, we find that when assessing the contribution of N-rich stars to the stellar halo field using either the TRI or TRI-CUT-DISC profiles as the best fitting N-rich star model, we obtain the same results as for the AXI profile, within the uncertainties.

The resulting density profiles for the halo (black) and N-rich star (orange) samples are displayed as a function of spherical radius in Fig. \ref{fig_dens}. It is clear that, while the halo density appears to decrease at a slower rate with increasing radius, the N-rich star sample exhibits a much greater decrease at high radii, and a much higher density within r $\lesssim$ 3 kpc, in comparison to the halo field population.
\subsection{Contribution of dissolved/evaporated Globular Clusters to the stellar halo of the Milky Way}
\label{results_mass}

Under the assumption that N-rich stars result from GC dissolution, it is interesting to estimate the total stellar mass contained in dissolved GC stars, in order to assess their total contribution to the total stellar-halo mass budget. Besides an estimate of the contribution to the stellar-halo budget by stars that once belonged to GCs, another output from our method is the spatial distribution of those stars. In this section, we estimate the mass density contributed from the halo and N-rich samples respectively as given by our best fitting density models as a function of distance from the Galactic Centre. Then we estimate the ratio between the mass in N-rich stars and the total stellar-halo mass. To determine the total mass contributed by field stars from GC origin to the stellar halo, we assume the ratio of "first population" (FP) to "second population" (SP) stars in GC to be 2 SP stars for every 1 FP star, adopting the minimal scenario proposed by \citet{Schiavon2017}. Although other FP-to-SP GC star ratios have been proposed (\citealp[e.g.][]{Bastian2015,Cabrera2015}), we chose to focus solely on the "minimal scenario" as it has been shown by \citet{Schiavon2017} that other scenarios can be excluded due to the predicted total GC star number counts exceeding the expected number of total halo stars. We then take the ratio of the disrupted GC mass with halo mass, to assess the mass contribution of the former to the halo field as a function of Galactocentric distance. Performing this exercise will not only lead to a deeper understanding of the origin of disrupted GC stars (and in consequence N-rich stars), but it will also provide a more clear understanding of the mass contribution from GC destruction to the total stellar halo mass budget.

We determine the mass in spherical annuli for both N-rich and halo stars, which in turn enables us to determine the mass from each sample as a function of spherical radius, as well as the ratio between the samples. As mentioned above, to estimate the mass contribution by stars that once belonged to GCs, we need to account for the contribution by former GC stars whose abundance patterns do not differ from that of the field population at same metallicity, which can be accounted for by assuming the FP-SP GC star ratio predicted from the minimal scenario. 
\begin{figure}
    \centering
    \includegraphics[width=\columnwidth]{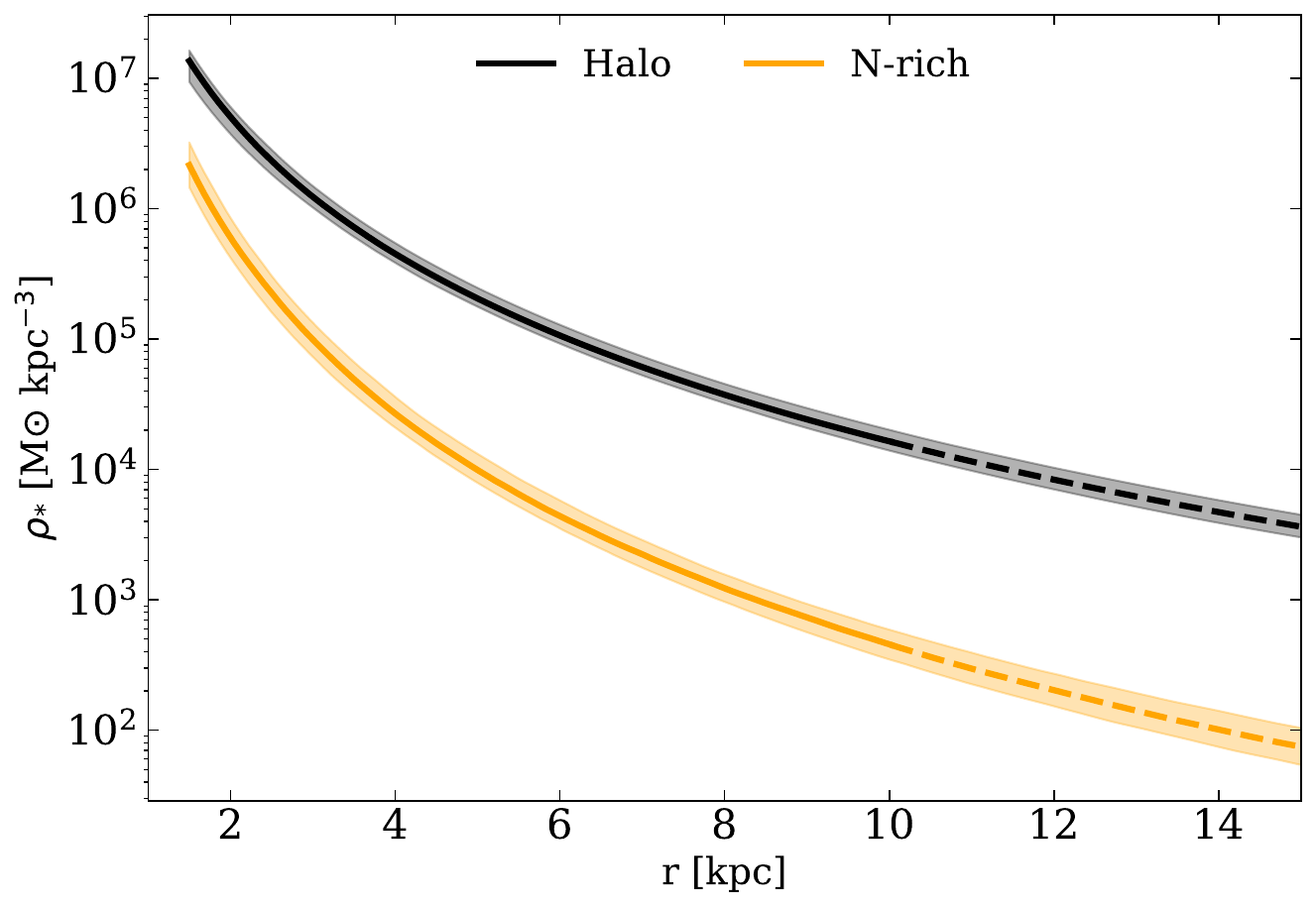}
    \caption{Integrated density as a function of spherical radius for the best fitting density profiles of the halo (black) and the N-rich (orange) star samples. The shaded regions mark the 16$^{th}$ and 84$^{th}$ percentile uncertainties. The dashed line indicates the region where, due to low sample numbers, the density is not strongly constrained.
    Both the halo and N-rich density profile follow a similar pattern within the outer r $\gtrsim$ 3 kpc region, however the N-rich density decreases more steeply.}
    \label{fig_dens}
\end{figure}

\begin{figure}
    \centering
    \includegraphics[width=\columnwidth]{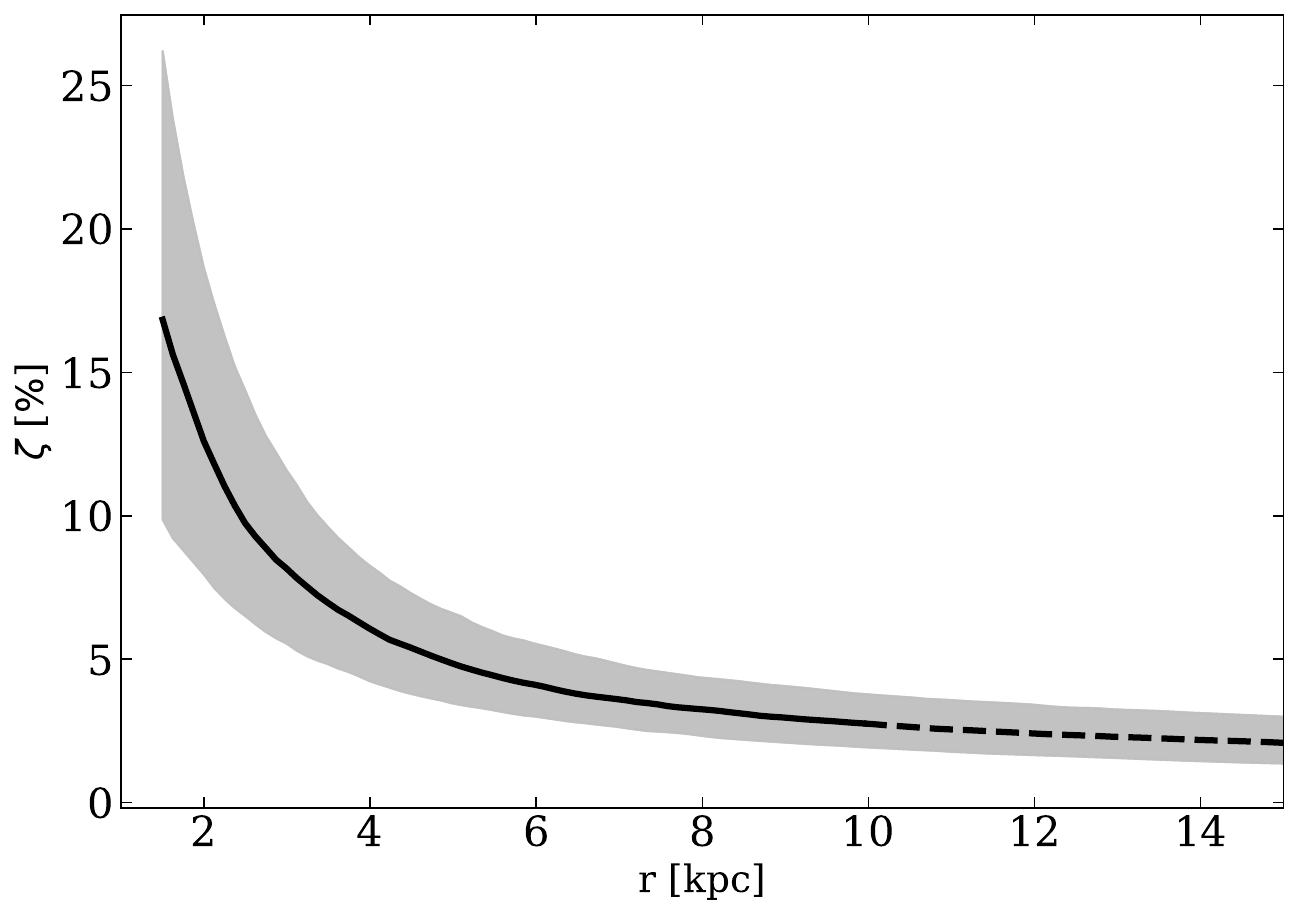}
    \caption{Mass density percentage ratio of N-rich stars and halo field stars as a function of spherical radius. The black solid line signifies the median value, while the shaded regions show the 16$^{th}$ and 84$^{th}$ percentile uncertainties. The dashed line indicates the    Galactocentric distance range where the density is not well constrained due to low numbers of N-rich stars. The mass density percentage ratio drops from $\zeta$ = 16.8$^{+10.0}_{-7.0}$$\%$ at r = 1.5 kpc to a value of $\zeta$ = 2.7$^{+1.0}_{-0.8}$$\%$ at r = 10 kpc. Under the minimal scenario assumption, one can multiply $\zeta$ by 1.5, and subtract the FP stars from the halo field, to obtain the total contribution from disrupted GC stars to the stellar halo.}
    \label{fig_massratio}
\end{figure}

Now that we have determined the density of both N-rich and halo stars as a function of Galactocentric distance, we focus our attention on estimating the ratio between N-rich stars and the total stellar halo. The results from this ratio estimation are shown as the black solid line in Fig. \ref{fig_massratio}, where the N-rich to halo ratio, hereafter denoted as $\zeta$, is given as a percentage. The shaded grey regions signify the 16$^{th}$ and 84$^{th}$ percentiles. In a similar fashion to Fig. \ref{fig_dens}, we choose to compute $\zeta$ between a 1.5 < r < 15 kpc range, for which is covered relatively well by the data. Our results show that $\zeta$ increases rapidly with decreasing radius, growing from a value of $\zeta$ = 2.7$^{+1.0}_{-0.8}$$\%$ at r = 10 kpc to a value of $\zeta$ = 16.8$^{+10.0}_{-7.0}$$\%$ at r = 1.5 kpc. Moreover, around the solar neighbourhood (i.e., r = 8 kpc), we find the ratio to be $\zeta$ = 3.3$^{+1.1}_{-1.0}$$\%$. Therefore, our estimates reveal approximately an eight-fold increase of the N-rich star to halo contribution in the inner Galactocentric regions, in agreement with previous findings \citep{Schiavon2017}.

Under the assumption of the minimal scenario presented in \citet{Schiavon2017}, whereby the ratio of FP to SP is 1-to-2, we estimate the ratio of dissolved GC stars to halo field stars to be $\zeta_{\mathrm{tot}}$ = 27.5$^{+15.4}_{-11.5}$$\%$ at r = 1.5 kpc. Along the same lines, we obtain an estimate of $\zeta_{\mathrm{tot}}$ = 4.2$^{+1.5}_{-1.3}$$\%$ at r = 10 kpc. Our results show that when accounting for selection effects in the observational data, the contribution of stars arising from GC dissolution and/or evaporation to the total stellar halo field is greater by a factor of$\sim$7-9 in the inner few kiloparsecs when compared to the outer regions of the Galaxy. Furthermore, we report that when repeating the methodology employing another well fitting density profile for the N-rich stars (in this case the TRI and TRI-CUT-DISC models), we find that, within the uncertainties, our results remain unchanged, thus validating both our estimates and procedure.

Now that we have shown the mass fraction contribution of stars arising from GC disruption to the halo field of the Galaxy as a function Galactocentric distance, we can compute an estimation of the mass contributed from GC escapees to any given volume or spherical shell. We chose to estimate the mass contributed by stars arising from GC destruction, under the minimal scenario assumption, within a volume spanning a radius between 1.5 and 3 kpc from the Galacic centre. For this shell volume, we obtain a total mass from stars arising from GC dissolution of M$_{\mathrm{GC,inner}}$ = 5.6$^{+2.8}_{-1.8}$ $\times$ 10$^{7}$ M$\odot$. Similarly, we compute the total mass of stars arising from GC disruption within a shell of$\sim$13.5 kpc in radius (from 1.5 kpc to 15 kpc), and find a mass estimate of M$_{\mathrm{GC,total}}$ = 9.6$^{+4.0}_{-2.6}$ $\times$ 10$^{7}$ M$\odot$. Thus, our results show that disrupted GC stars contribute a significant amount of mass to the stellar halo of the Galaxy. This notable mass contribution is observed at all Galactocentric scales, however is more dominant at smaller radii, as shown in Fig. \ref{fig_massratio}. 

\section{Discussion}
\label{discussion}
The results obtained in Fig. \ref{fig_massratio} suggest that the fraction of N-rich star mass as a function of halo field mass is much greater in the inner regions of the Galaxy when compared to the outer regions. In agreement with observational estimates (\citealp{Martell2010,Carollo2013,Martell2016,Schiavon2017,Trincado2019,Koch2019,Lee2019}), and theoretical predictions (\citealp{Tremaine1975,Gnedin2014,Reina2019,Hughes2020}), we hypothesise that N-rich stars result from the destruction of pre-existing GCs. Assuming this hypothesis is correct, our findings may have important repercussions for the current understanding of the formation and evolution of the Galactic GC system, the presence of multiple stellar populations in GCs, possibly the formation and evolution of the Milky Way bulge, and even the mass assembly history of the Galaxy. In the following subsections, we discuss and compare the results from our stellar halo density modelling and mass estimation to previous work.

\subsection{Comparison with previous halo density work}
\label{discussion_densmodelling}

In this section, we perform a detailed comparison of our findings with those from previous works. The contrasting method employed in this work allows for an interesting comparison, one that will help shed light into the nature of the density of the stellar halo of the Galaxy.

The majority of early work focused on modelling the density of the Galactic stellar halo employed different stellar types as tracers of the halo population, including either Main Sequence Turn Off (MSTO) stars (\citealp{Morrison2000,Juric2008}) or Blue Horizontal branch (BHB) stars \citep{Yanny2000}, for which the photometry is relatively easily calibrated to provide accurate distances \footnote[5]{Listed in this subsection are only a few of the more recent examples of an extensive list of studies focused on estimating the density of the stellar halo. For a more comprehensive account of previous work see \citet{Helmi2008}.}. By fitting single power-law models of the form:
\begin{equation}
\rho(r_{\mathrm{e}}) = r_{\mathrm{e}}^{-\alpha},
\end{equation}
these groups found an exponent of $\alpha$$\sim$3 and find a flattening requirement (given by Eq \ref{eq2} when equating $p$ = 1) to fit star counts well. A pivotal work by \citet{Deason2011} showed that BHB stars in SDSS could be best fit by introducing a break radius at $r_{\mathrm{b}}$$\sim$27 kpc in their exponential profile, thus fitting a broken exponential, for which the inner slope is shallow $\alpha_{\mathrm{in}}$$\sim$2.3, and the outer slope is steep, $\alpha_{\mathrm{out}}$$\sim$4.6. Along those lines, \citet{Whitten2019} used a BHB sample from a cross-match of Pan-STARRS and GALEX data, and studied their relative age-distribution across the Galactic halo. The results from this work showed that the BHB relative age-distribution of the Milky Way is also best modelled by a broken profile, finding an older population within the ($R_{\mathrm{GC}}$$\sim$14 kpc) break radius. Another crucial study by \citet{Xue2015} used SDSS-SEGUE giants to show that such a profile can be accommodated with a single power law of steep exponent $\alpha$$\sim$4.2, with flattening parameter \textit{q} varying as a function of Galactocentric distance between \textit{q}$\sim$0.5 and 0.8. Later, \citet{Iorio2018} found that a similar profile provided a good fit to a sample of RR Lyrae stars within a Galactocentric distance of$\sim$30 kpc, such that $\alpha$$\sim$2.96 with 0.57 < \textit{q} < 0.75. A recent study found that the APOGEE DR14 giant data were best fit by a similar single power-law profile, however with the inclusion of a disc and cut-off term \citep{Mackereth2019b}, such as the model employed to fit the halo sample in this study. The power law slope obtained for this model was found to be steep $\alpha$$\sim$3.5, presenting a flattened triaxial ellipsoid, \textit{p}$\sim$0.73 and \textit{q}$\sim$0.56, slightly rotated in the \textit{Z} and \textit{Y} axes.

The behaviour resulting from the fit to the APOGEE data is similar to work from the literature. Fitting a likely halo sample of stars with metallicities between --2.5 < [Fe/H] < --1, we find the data are well matched by a triaxial profile with a moderately steep exponent $\alpha$$\sim$3.5, flattened with \textit{p}$\sim$0.8 and \textit{q}$\sim$0.5. We also find the triaxial ellipsoid to be minorly rotated to the $Z$ and $Y$ axes, by $\theta$$\sim$3-4 degrees and $\phi$$\sim$14 degrees, respectively. Finally,
since our data is all contained within a $\sim$20 kpc radius, the issue of flattening beyond the break radius found in other studies is irrelevant to this work. Thus, we can conclude that the resulting profile and its associated parameters values are consistent with the results found in previous studies.

\subsection{The contribution of GCs to the stellar halo of the Galaxy}
\label{discussion_mass}
\subsubsection{Mass ratio of stars arising from GC dissolution and/or evaporation}
\label{discussion_ratio}
Employing the method described in Section \ref{method}, we have assessed the fraction of the halo stellar mass contributed by N-rich stars as a function of Galactocentric distance between 1.5 < r < 15 kpc. We have shown that N-rich stars contribute a $\zeta$ = 16.8$^{+10.0}_{-7.0}$$\%$ to the halo field at$\sim$1.5 kpc from the Galactic Centre (see Fig. \ref{fig_massratio}), resulting in a much larger contribution when compared to the outer regions of the Galaxy (i.e. r$\sim$10 kpc) for which we find a ratio of $\zeta$ = 2.7$^{+1.0}_{-0.8}$$\%$. Under the assumption of the minimal scenario proposed by \citet{Schiavon2017}, whereby the ratio of FP-to-SP GC star is assumed to be 2 SP stars for every 1 FP star, we estimate the total contribution from disrupted star clusters$\sim$1.5 kpc from the Galactic Centre to be $\zeta_{\mathrm{tot}}$ = 27.5$^{+15.4}_{-11.5}$$\%$. Along those lines, we estimate the contribution from GC dissolved stars to the stellar halo at r = 10 kpc to be $\zeta_{\mathrm{tot}}$ = 4.2$^{+1.5}_{-1.3}$$\%$. Within the uncertainties, our estimate for the contribution by stars that once belonged to GCs to the stellar halo mass budget at Galactocentric distances of r = 10 kpc is in agreement with the low bound of the 2-5$\%$ estimate from the literature for the halo of the Galaxy (\citealp{Martell2016,Koch2019,Reina2019}). Similarly, our estimate for the contribution at r = 1.5 kpc from the Galactic Centre is in agreement with the 19-24$\%$ estimate from \citet{Schiavon2017}, but is only partly in agreement with the theoretical predictions from \citet{Hughes2020}. Specifically, within the uncertainties, our estimate falls in the upper range of the predictions provided by the E-MOSAICS suite of simulations, which predict a contribution ranging between 0.3 and 14\% for Milky Way analogs, depending on a galaxy's accretion history. 

Interestingly, the largest fractional contribution to the stellar mass budget by disrupted globular clusters is attained in the simulations by \citet{Hughes2020} by galaxies that underwent a phase of intense accretion in their early lives.  \citet{Mackereth2018}, analysing EAGLE simulations (\citealp[][]{Schaye2015,Crain2015}) established a link between those types of accretion histories with the (rare) presence of an $\alpha$ bimodality in the simulated galaxies' disk populations.  According to \citet{Hughes2020}, the physical basis for this connection is the high gas pressure brought about by intense accretion which, on one hand lowers the gas consumption timescale leading up to formation of an $\alpha$-enhanced population, and on the other hand creates an environment that is hostile to globular cluster survival.  

In this context, it is important to notice that Kisku et al. (2020, in prep.) studied the abundance patterns of N-rich stars in the inner Galaxy, suggesting that roughly 1/2 of the N-rich stars with [Fe/H] < -1 were actually accreted, and were likely associated with the progenitor of the Inner Galaxy Structure (i.e., IGS), identified by \citet{Horta2020b}.  

\subsubsection{Mass in dissolved GCs}
As mentioned in Section \ref{results_mass}, our methodology enables us to determine the mass for both the stellar halo and for N-rich stars within a sphere of any given radius. Under the minimal scenario assumption (see Section \ref{results_mass}), we estimate a total stellar mass arising from GC remnants, within a shell ranging from 1.5 -- 15 kpc in radius around the Galactic Centre, to be M$_{\mathrm{GC,total}}$ = 9.6$^{+4.0}_{-2.6}$ $\times$ 10$^{7}$ M$\odot$. Our estimates confirm theoretical predictions by different groups (\citealp[$\sim$10$^{7}$ - $10^{8}$ M$\odot$,][]{Tremaine1975,Gnedin2014}). Furthermore, we find that the estimated mass from stars arising from GC disruption and/or evaporation to be approximately$\sim$3-4 times greater than the total mass in all existing Galactic GCs \citep[$\sim$2.8 $\times$ 10$^{7}$ M$\odot$,][]{Kruijssen_Zwart2009}.
This result would imply that the Galactic GC system was initially four to five times larger, where approximately only one fourth/fifth survived, resulting in the$\sim$150 GCs observed today. However, due to the metallicity cuts employed in this study, our estimated mass is biased low, since it does not include the mass from stars arising from disruption and/or evaporation of GCs with [Fe/H] > --1. Moreover, our estimate only accounts for GCs that were massive enough to develop multiple populations (MP).  It is likely that GCs were formed with masses below that threshold that nevertheless were destroyed, contributing only FP stars to the field population.  Those of course are not accounted for by our chemical tagging, so that our estimated contribution of GCs to the stellar mass budget of halo field populations should be taken as a lower limit.

By the same token, since our results show that there is a much higher contribution of dissolved GC stars in the inner galaxy, we estimated the total stellar mass arising from GC disruption, within a hollow sphere of 1.5 - 3 kpc in radius around the Galactic Centre, to be $M_{\mathrm{GC,inner}}$ = 5.6$^{+2.8}_{-1.8}$ $\times$ 10$^{7}$ M$\odot$. We find our estimate to be greater than the predicted$\sim$10$^{7}$ M$\odot$ from \citet{Gnedin2014}. Similarly, within the uncertainties, we find our result to be slightly smaller than the estimate given in \citet{Schiavon2017}, who found that the mass in destroyed GCs within 2 kpc to be a factor of 6-8 higher than the mass of existing GCs.  The difference is due to three factors:  (1) the different volumes included in the calculation; (2) the different models for the inner halo adopted, and (3) the different metallicity ranges considered.  

\subsection{Supporting evidence for the GC origin of N-rich stars}
Despite the growing evidence for the GC origin of N-rich stars, there have been many alternate scenarios proposed for the formation of such population. Such alternative formation channels range from the notion that N-rich stars are the oldest stellar population in the Galaxy which formed in high density environments \citep{Bekki2019}, to the idea that N-rich stars were formed in the same molecular clouds that GCs were formed in, however were never gravitationally bound to them. In this subsection, we discuss the results from previous studies which have shown, using measurements other than chemical compositions, that N-rich stars are likely formed from the dissolution and/or evaporation of GCs. 

The availability of the 6D phase space information provided by the \emph{Gaia} survey has made possible the estimation of the integrals of motion (hereafter IoM) of stars in the Milky Way. Such properties are adiabatic invariants, and thus retain some information about the origin of their parent population. A recent study by \citet{Savino2019} determined the IoM for a sample of 65 CN-strong stars, from the work by \citet{Martell2010}, as well as for the Galactic GC system. In their study, a direct comparison of the IoM values, as well as the metallcity values, was performed for every N-rich-GC pair, associating a likelihood of these being from the same distribution. The results from that study showed that a considerable fraction of CN-strong stars display the same IoM values as existing GCs, thus supporting the notion that N-rich stars arise from GC dissolution and/or evaporation.

A similar chemo-dynamical analysis of halo stars and Galactic GCs was performed by \citet{Hanke2020}. In that study, three separate methodologies were employed to attempt to link halo stars to existing GCs. These ranged from statistically linking halo stars positioned nearby existing Galactic GCs, to linking CN-strong stars (i.e. N-rich stars) kinematically to existing GCs, to attempting to find halo stars which displayed similar kinematics to those CN-strong stars. The authors showed that over 60$\%$ of their N-rich star population presented IoM which could be statistically linked to known GCs, and that around$\sim$150 halo stars could be associated with a GC origin.
Separately, a study by \citet{Tang2020} performed a comparison between the IoM of$\sim$100 CN-strong (metal poor) stars from the LAMOST DR5 data set with metal poor halo stars. The authors from this study concluded that the CN-strong stars do not present similar kinematics to the halo field population, but resembled that of the inner halo where there is a high density of GCs with similar chemical compositions. Based on their kinematic results, the authors supported a GC origin for the CN-strong stars. In a final remark, we find the findings of \citet{Tang2020} to be in agreement with the results from \citet{Carollo2013}. Using an independent sample CN-strong stars, these authors found that these follow the velocity distribution of the "inner halo population" (IHP, as defined in their work). 

The kinematic results from the aforementioned work, when coupled with the results from chemical composition studies (\citealp[e.g.,][]{Martell2010,Martell2016,Schiavon2017,Koch2019,Lee2019}), support the hypothesis that N-rich stars arise from the disruption and/or evaporation of GCs.

\section{Summary and Conclusions}
\label{conclusions}
In this paper, we report a study of the spatial variation of the density of N-rich and normal halo field stars.  In this way we assessed the contribution of N-rich stars, and by inference that of dissolved GCs, to the stellar mass budget of the Galactic halo, as well as its variation as a function of Galactocentric distance. A summary of our results is listed as follows:\\

\begin{itemize}

  \item We identified in a parent sample of 1455 halo stars with --2.5 < [Fe/H] < --1, using a Gaussian Mixture Model, 46 N-rich stars that are distributed throughout the Galaxy (see Fig. \ref{fig_gmm} and Fig. \ref{fig_spatial}), and are not bound to existing GCs. The N-rich stars present a N-C abundance anti-correlation and an N-Al abundance correlation (see Fig. \ref{fig_cfes} and Fig. \ref{fig_alfes}, respectively), and are likely second population stars that once were bound to a GC (which may or may not still exist) and now reside in the halo field of the Galaxy.

  \item We show that once the survey selection effects are accounted for (see Appendix A in \citealp{Mackereth2019} for details), the halo APOGEE data between --2.5 < [Fe/H] < --1 are well fit by a triaxial single power law with exponent $\alpha$$\sim$3.5, and flattening parameters $p\sim$0.8, $q\sim$0.5, with major axis only slightly misaligned with the axis connecting the Sun and the Galactic centre.

  \item Similarly, we show that the APOGEE  N-rich star data between --2.5 < [Fe/H] < --1 are well fit by an axisymmetric profile, defined by a single power law of slope $\alpha$$\sim$4.5. The best fitting model is flattened along the \emph{Z} axis, with a flattening value similar to that of the halo sample \textit{q}$\sim$0.5 (see Fig. \ref{fig_cornernrich}).

\item We find a contribution of N-rich stars to the stellar halo of $\zeta$ = 16.8$^{+10.0}_{-7.0}$$\%$ at r = 1.5 kpc. However, this ratio drops by a factor of $\sim$6 at large Galactocentric distances (r = 10 kpc), where we estimate a contribution of $\zeta$ = 2.7$^{+1.0}_{-0.8}$$\%$.
\item Assuming that the ratio of first population-to-second population stars in GCs is 1-to-2, we find that stars arising from GC disruption contribute $\zeta_{\mathrm{tot}}$ = 27.5$^{+15.4}_{-11.5}$$\%$ to the mass of the stellar halo at$\sim$1.5 kpc from the Galactic Centre. Conversely, the contribution of GC escapees at larger Galactocentric distances (i.e. r = 10 kpc) is lower, namely $\zeta_{\mathrm{tot}}$ = 4.2$^{+1.5}_{-1.3}$$\%$. Such estimates are in agreement (within the uncertainties) with previous estimates from the literature for the inner Galaxy \citep[e.g.,][]{Schiavon2017} and the outer Galactic halo (\citealp[e.g.,][]{Martell2016,Koch2019}).

\item We integrate the total stellar mass between --2.5 < [Fe/H] < --1 and 1.5 < r < 15 kpc using Eq \ref{eq8} and our best-fitting halo profile, and estimate the mass of the stellar halo to be M$_{\mathrm{*,halo}}$ = 8.3 $^{+1.5}_{-1.3}$ $\times$ 10 $^{8}$ M$\odot$.

\item We integrate the total stellar mass between --2.5 < [Fe/H] < --1 and 1.5 < r < 15 kpc using Eq \ref{eq8} and our best-fitting N-rich model, and estimate the mass from N-rich stars to be M$_{\mathrm{*,N-rich}}$ = 6.4 $^{+2.6}_{-1.7}$ $\times$ 10 $^{7}$ M$\odot$.

\item Using the same technique, and under the assumption of a 1:2 first:second population GC ratio, we estimate the total stellar mass between --2.5 < [Fe/H] < --1 and 1.5 < r < 15 kpc arising from GC dissolution and/or evaporation to be $M_{\mathrm{GC,total}}$ = 9.6 $^{+4.0}_{-2.6}$ $\times$ 10$^{7}$ M$\odot$. For a spherical volume ranging from 1.5-3 kpc, we obtain an estimated mass of $M_{\mathrm{GC,inner}}$ = 5.6$^{+2.8}_{-1.8}$ $\times$ 10$^{7}$ M$\odot$. This total dissolved/evaporated GC mass is approximately 3-4 times greater than the total mass in all existing Galactic GCs \citep[namely, $\sim$2.8$\times$10$^{7}$ M$\odot$][]{Kruijssen_Zwart2009}.

\item We speculate that the increased contribution of GC dissolution towards the inner regions of the Galaxy may be associated with enhanced merger activity in the early life of the Milky Way.  Some of these merging systems (e.g., IGS \citealp{Horta2020b}) may have brought with them a population of extragalactic N-rich stars (Kisku et al, in prep.).  In addition, the enhanced merging activity in the early life of the Milky Way may also have given rise to conditions that led to efficient {\it in situ} formation and destruction of GCs, leaving behind a large population of N-rich stars in the inner Galaxy field.
\end{itemize}

In this paper, we have mapped the spatial distribution of N-rich stars in the Galactic halo, determining their contribution to the stellar mass budget as a function of Galactocentric distance. 

Our study of the contribution of N-rich stars, commonly recognized as SP GC stars residing in the field, to the halo field constrains the mass contribution from GC disruption and/or evaporation to the total mass of the stellar halo, but also provides insights into the mass assembly history of the Milky Way. The order of magnitude increase in halo samples afforded by upcoming surveys such as WEAVE \citep{Dalton2014} and 4-MOST \citep{deJong2019} will place key constraints on our understanding of the mass assembly of the Milky Way halo, and the role played by globular clusters in this process.

\section*{Acknowledgements}
We thank the many people around the world whose hard work fighting the ongoing COVID-19 pandemic has made it possible for us to remain safe and healthy for the past several months. The authors thank Nate Bastian, Diederik Kruijssen, Meghan Hughes, and Alvio Renzini for helpful discussions and/or comments on an early version of this manuscript. DH thanks Sue, Alex and Debs for their moral encouragement, and acknowledges an STFC doctoral studentship. JTM acknowledges support from the ERC Consolidator Grant funding scheme (project ASTEROCHRONOMETRY, G.A. n. 772293), the Banting Postdoctoral Fellowship programme administered by the Government of Canada, and a CITA/Dunlap Institute fellowship. The Dunlap Institute is funded through an endowment established by the David Dunlap family and the University of Toronto. TCB acknowledges partial support for this work from grant PHY 14-30152; Physics Frontier Center/JINA Center for the Evolution of the Elements (JINA-CEE), awarded by the US National Science Foundation. DMN acknowledges support from NASA under award Number 80NSSC19K0589. ARL acknowledges financial support provided in Chile by Comisi\'on Nacional de Investigaci\'on Cient\'ifica y Tecnol\'ogica (COINICYT) through the FONDECYT project 1170476 and by the QUIMAL project 130001. CAP is thankful to the spanish goverment for research funding (AYA2017-86389-P). DAGH acknowledges support from the State Research Agency (AEI) of the
Spanish Ministry of Science, Innovation and Universities (MCIU) and the
European Regional Development Fund (FEDER) under grant AYA2017-88254-P.

Funding for the Sloan Digital Sky Survey IV has been provided by
the Alfred P. Sloan Foundation, the U.S. Department of Energy Office
of Science, and the Participating Institutions. SDSS acknowledges
support and resources from the Center for High-Performance Computing
at the University of Utah. The SDSS web site is www.sdss.org. SDSS
is managed by the Astrophysical Research Consortium for the
Participating Institutions of the SDSS Collaboration including the
Brazilian Participation Group, the Carnegie Institution for Science,
Carnegie Mellon University, the Chilean Participation Group, the
French Participation Group, Harvard-Smithsonian Center for Astrophysics,
Instituto de Astrof\'{i}sica de Canarias, The Johns Hopkins University,
Kavli Institute for the Physics and Mathematics of the Universe
(IPMU) / University of Tokyo, the Korean Participation Group,
Lawrence Berkeley National Laboratory, Leibniz Institut f\"{u}r Astrophysik
Potsdam (AIP), Max-Planck-Institut f\"{u}r Astronomie (MPIA Heidelberg),
Max-Planck-Institut f\"{u}r Astrophysik (MPA Garching), Max-Planck-Institut
f\"{u}r Extraterrestrische Physik (MPE), National Astronomical Observatories
of China, New Mexico State University, New York University, University
of Notre Dame, Observat\'{o}rio Nacional / MCTI, The Ohio State University,
Pennsylvania State University, Shanghai Astronomical Observatory,
United Kingdom Participation Group, Universidad Nacional Aut\'{o}noma
de M\'{e}xico, University of Arizona, University of Colorado Boulder,
University of Oxford, University of Portsmouth, University of Utah,
University of Virginia, University of Washington, University of
Wisconsin, Vanderbilt University, and Yale University.

{\it Software:} NumPy \citep{NumPy}, Matplotlib \citep{Hunter:2007}.
\section*{Data availability}
All APOGEE DR16 data used in this study is publicly available and can be found at: https: //www.sdss.org/dr16/




\bibliographystyle{mnras}
\bibliography{refs}


\appendix

\section{Fitting the N-rich star sample}
\label{fitting_Nrich}
We set out to determine the best fit model to the sample of N-rich stars. Our methodology consists on fitting several different stellar halo density models and determining their logarithmic maximum likelihood (ln($\mathcal{L}_{\mathrm{\mathrm{max}}}$)) and Bayesian Information Criterion \citep[BIC, ][]{Schwarz1978} values (see Section \ref{bic}). We then compare the values obtained for each density model, and take the profile with the lowest BIC value to be our best fit model. If two models obtain the same ln($\mathcal{L}_{\mathrm{\mathrm{max}}}$) value, the BIC value gives preference to the model with the lowest number of free ranging parameters, therefore choosing the simplest model. 

\begin{figure*}
    \centering
    \includegraphics[width=\textwidth]{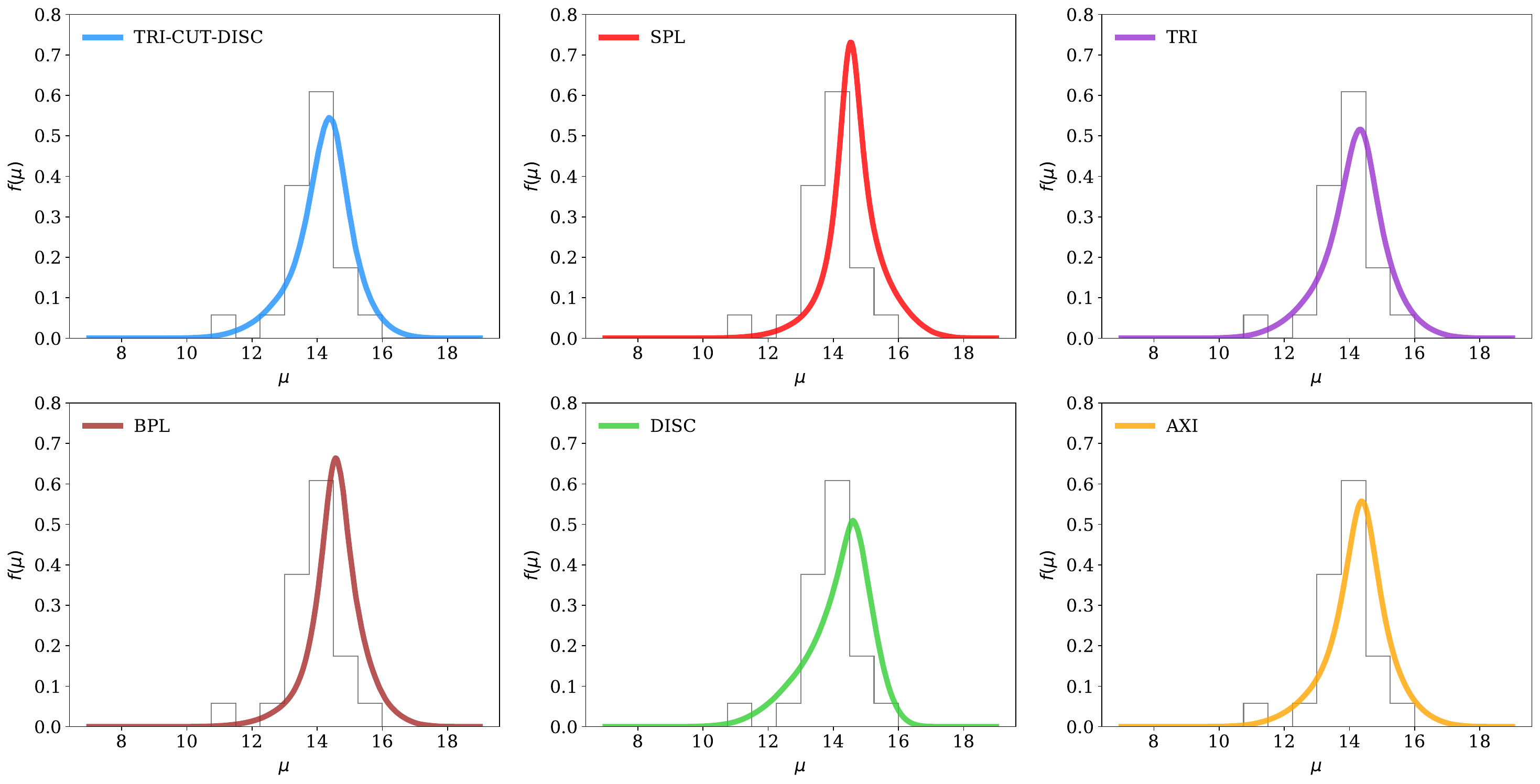}
    \caption{Density profile fits to the N-rich star data from Section \ref{data}. Each profile is obtained by taking the median posterior parameter value obtained from the 10,000 MCMC samples. The AXI profile is the best-fitting profile, closely followed by the TRI and TRI-CUT-DISC profile.}
    \label{fig_nrich_fits}
\end{figure*}

As described in Section \ref{n-rich_models}, our N-rich sample is fit by a spherical single power law (SPL), an axisymmetric single power law (AXI), a triaxial power law (TRI), a rotated triaxial power law with the inclusion of a cut-off term and a disc contamination parameter (TRI-CUT-DISC), a broken power law (BPL), and an exponential disc profile (DISC). The resulting BIC and ln($\mathcal{L}_{\mathrm{\mathrm{max}}}$) values obtained using the median posterior parameter of the 10,000 MCMC realisations for each model are listed in Table \ref{tab1}. From our results, we find that the best fit model to our N-rich sample is the AXI model, and therefore choose to use this model for the remainder of the analysis. However, as is apparent from Fig. \ref{fig_nrich_fits}, the TRI and TRI-CUT-DISC density profiles also provide good fits to the N-rich star data. Moreover, we repeat the methodology for determining the percentage ratio contribution of N-rich stars to the halo field employed in the main body of the paper using the TRI and TRI-CUT-DISC profile as the best-fitting N-rich models, in order to check if our results vary when adopting a different model for the N-rich star sample. The results from this comparison show that our initial findings remain unchanged, and suggest that the N-rich star sample can either be well modelled by the AXI, TRI, and TRI-CUT-DISC profiles.\\

\setlength{\tabcolsep}{18pt}
\begin{table*}
\begin{tabular}{ |p{2cm}|p{12.cm}|  }
\hline
Density profile & Best-fit parameters\\
\hline
\hline
  SPL & $\alpha$=3.86$^{+0.21}_{-0.18}$\\
  \\
  \textbf{AXI} &\textbf{$\alpha$=4.47$^{+0.23}_{-0.22}$, $q$=0.47$^{+0.05}_{-0.04}$}\\
  \\
  TRI& $\alpha$=4.54$^{+0.26}_{-0.24}$, $p$=0.85$^{+0.09}_{-0.10}$, $q$=0.44$^{+0.05}_{-0.05}$ \\
  \\
  TRI-CUT-DISC&$\alpha$=4.30$^{+0.12}_{-0.31}$ , $\beta$=0.03$^{+0.03}_{-0.02}$, $p$=0.60$^{+0.21}_{-0.09}$, $q$=0.41$^{+0.11}_{-0.07}$, $\eta$=0.20$^{+0.03}_{-0.04}$, $\theta$=0.06$^{+0.2}_{-0.06}$ [$^{\circ}$], $\phi$=0.25$^{+0.07}_{-2.38}$ [$^{\circ}$],
  $f_{\mathrm{disc}}$=0.04$^{+0.11}_{-0.03}$\\  
  \\
  BPL& $\alpha_{\mathrm{in}}$=3.19$^{+0.52}_{-0.46}$, $\alpha_{\mathrm{out}}$=6.05$^{+1.25}_{-1.37}$, $R_{\mathrm{break}}$=8.13$^{+2.03}_{-1.82}$[kpc] \\
  \\
  DISC& 1/$h_{\mathrm{R}}$=0.41$^{+0.04}_{-0.04}$[kpc$^{-1}$],1/$h_{\mathrm{Z}}$=0.64$^{+0.09}_{-0.08}$[kpc$^{-1}$]\\
  \hline
\hline
\end{tabular}
\caption{Resulting best-fit parameters for the different functional density forms fitted to the N-rich star sample. The best-fitting model (AXI) is highlighted in bold.}\label{tab2}
\end{table*}

\bsp	
\label{lastpage}
\end{document}